# Explainable Dual-Attention Tabular Transformer for Soil Electrical Resistivity Prediction: A Decision Support Framework for High-Voltage Substation Construction

**Warat Kongkitkul, Sompote Youwai* and Warut Sakulpojworachai**

AI Research Group, Department of Civil Engineering, King Mongkut's University of technology Thonburi
*Corresponding Author Email: sompote.you@kmutt.ac.th

**Abstract**
This research introduces a novel dual-attention transformer architecture for predicting soil electrical resistivity, a critical parameter for high-voltage substation construction. Our model employs attention mechanisms operating across both features and data batches, enhanced by feature embedding layers that project inputs into higher-dimensional spaces. We implements Particle Swarm Optimization for hyperparameter tuning, systematically optimizing embedding dimensions, attention heads, and neural network architecture. The proposed architecture achieves superior predictive performance (Mean Absolute Percentage Error: 0.63%) compared to recent state of the art models for tabular data. Crucially, our model maintains explainability through SHapley Additive exPlanations value analysis, revealing that fine particle content and dry density are the most influential parameters affecting soil resistivity. We developes a web-based application implementing this model to provide engineers with an accessible decision support framework that bridges geotechnical and electrical engineering requirements for the Electricity Generating Authority of Thailand. This integrated approach satisfies both structural stability and electrical safety standards, improving construction efficiency and safety compliance in high-voltage infrastructure implementation.

## 1. Introduction

Soil electrical resistivity represents a fundamental parameter in geotechnical and electrical engineering, significantly influencing the design efficacy of earthing systems in high-voltage infrastructure. As power transmission networks expand globally, precise soil resistivity assessment has become increasingly critical for ensuring both operational reliability and safety compliance. The Electricity Generating Authority of Thailand (EGAT) exemplifies organizations requiring accurate resistivity predictions for high-voltage substation construction, where effective grounding directly impacts system protection against fault currents. According to EGAT standards, backfill soils in substations must not exceed certain value to meet criteria for main substation establishment (Sangprasat et al., 2024). Therefore, predicting soil resistivity is essential for determining optimal substation locations, understanding soil properties, implementing appropriate soil improvements, and preventing degradation in areas of concern regarding soil resistivity values.

Current substation design methodologies exhibit a significant gap between geotechnical engineering requirements and electrical system performance criteria. This fundamental disconnect creates substantial challenges for engineers attempting to optimize both structural integrity and electrical efficiency in substation installations. The relationship between soil resistivity and geotechnical properties presents a complex challenge that has attracted significant research attention. Traditional approaches have relied on empirical formulas derived from laboratory experiments, often yielding relationships that are specific to particular soil types or environmental conditions. While these methods have provided valuable insights, they frequently struggle to capture the intricate non-linear interactions between multiple soil parameters. Previous studies by (Bai et al., 2013) and (Cardoso and Dias, 2017) have demonstrated that soil resistivity exhibits sophisticated dependencies on various factors, including moisture content, dry density, void ratio, and mineral composition. Recent research by Sangprasat et al. (2024) confirms that soil resistivity is highly dependent on water content, revealing a significant inverse relationship between soil resistivity and water content and establishing water content as a dominant factor influencing resistivity values. Despite these advancements in understanding, current prediction methods continue to face substantial limitations. Empirical models often oversimplify these relationships, while sophisticated numerical approaches may sacrifice physical interpretability for accuracy. Statistical regression techniques, while useful, typically require predefined functional forms that may not capture the true underlying physical relationships. Several researchers have attempted to address these limitations by applying Multi-Layer Perceptron (MLP) to predict the electrical properties of soil (Alsharari et al., 2020; Ozcep et al., n.d.); however, significant discrepancies persist between predicted values and ground truth data, with such models functioning as "black boxes" that lack explainability and transparent implementation for real-world applications. This opacity is further compounded by the tabular nature of soil resistivity data, which presents unique challenges—despite appearing amenable to capture or simulation, tabular data often yields higher prediction errors due to limited correlation between samples, unlike image data or natural language data which exhibit patterns or connections between adjacent features, thus making the establishment of robust predictive and simulation models for soil resistivity measurements a significant ongoing challenge in the field.

Traditional methods for tabular data prediction, as alternatives to Multi-Layer Perceptron (MLP), include tree-based algorithms such as XGBoost (Chen and Guestrin, 2016) and CatBoost (Prokhorenkova et al., 2018). These approaches have consistently demonstrated superior performance compared to MLP architectures in simulating tabular data. Their effectiveness stems from their inherent ability to capture non-linear relationships and interactions between features without requiring extensive preprocessing. The tree-based ensemble methodology enables these algorithms to handle missing values efficiently, manage high-dimensional data, and automatically identify relevant feature interactions through recursive partitioning techniques. Additionally, these methods provide feature importance metrics that enhance model interpretability, a critical factor in scientific applications where understanding variable contributions is essential. Recently, transformer architectures (Vaswani et al., 2017) have been employed for tabular data prediction due to their capacity to extract attention patterns between individual features and their versatility in handling both categorical and numerical data simultaneously (Cholakov and Kolev, 2022; Hollmann et al., 2025; Huang et al., 2020). The self-attention mechanism in transformers enables the model to learn complex dependencies within tabular structures by dynamically weighting the importance of different feature combinations across various contexts. This approach differs fundamentally from both neural networks and tree-based methods by allowing the model to consider all features simultaneously rather than sequentially or hierarchically. Transformer models can effectively capture long-range dependencies between features that might be missed by traditional methods, while also accommodating mixed data types through appropriate embedding techniques. Their adaptability to various data distributions and ability to scale with increasing feature dimensionality makes them particularly promising for complex tabular data prediction tasks in scientific domains where underlying relationships may not follow conventional patterns.

This research introduces Dual-Attention Tabular Transformer (DTT) as an innovative machine learning architecture for addressing the soil resistivity prediction challenge. Drawing inspiration from Hollman et al. (2025), the model implements attention mechanisms that operate not only across features but also within and between data batches, enabling the study of correlations at multiple levels. Unlike previous approaches, this research incorporates an embedding layer that projects features into higher-dimensional spaces, facilitating enhanced feature extraction through the transformer architecture. These innovations allow Explainable Tabular Transformers to leverage self-attention mechanisms specifically optimized for structured data, enabling them to process heterogeneous geotechnical parameters simultaneously while preserving feature interpretability and capturing complex non-linear relationships between soil classification metrics and electrical properties, all while providing explainable predictions through attention weight visualization and feature attribution techniques. Building upon this technical foundation and the experimental framework established by previous researchers, particularly the comprehensive soil testing methodologies developed at King Mongkut's University of Technology Thonburi (KMUTT), we apply our approach to analyze resistivity data from both lateritic soil and fine sand samples, combining rigorous experimental data collection using standardized four-electrode Wenner array configurations with advanced computational techniques to derive robust predictive models that integrate standardized Unified Soil Classification System (USCS) classification parameters with electrical resistivity measurements obtained through Wenner Four-Point Method testing according to IEEE 81 standards (IEEE Power and Energy Society, 2012).

The objectives of this investigation are threefold: (1) to develop an Explainable Dual-Attention Tabular Transformer that identifies accurate and interpretable relationships for soil resistivity prediction, (2) to validate these models using experimental data across diverse soil conditions, and (3) to elucidate the physical mechanisms governing soil electrical properties through model explainability. This research addresses the gap between empirical observations and theoretical understanding, providing quantitative tools for geotechnical engineers while advancing fundamental knowledge of soil behavior. The Explainable Tabular Transformer architecture incorporates interpretability mechanisms via SHAP (SHapley Additive exPlanations) values and attention weight visualization, quantifying the contribution of specific soil parameters to resistivity predictions. This explanatory capability resolves a significant limitation in current predictive models, enabling robust implementation in engineering specifications. By establishing quantifiable correlations between USCS classification and electrical resistivity, this work aims to develop an integrated specification framework that satisfies both geotechnical stability requirements and electrical safety standards, thereby enhancing construction efficiency, minimizing remediation costs, and improving safety compliance in high-voltage substation implementation.

The contribution of this research are as follows:

- We introduce a novel dual-attention transformer architecture specifically designed for soil resistivity prediction, which employs attention mechanisms operating across both features and data batches simultaneously.
- The model incorporates an embedding layer that projects features into higher-dimensional spaces before applying transformer attention mechanisms, improving feature extraction capabilities.
- The architecture provides explainability through SHapley Additive exPlanations values and attention weight visualization, making the relationships between soil parameters and resistivity predictions interpretable, unlike black box models.
- It demonstrates superior predictive performance compared to existing methods, achieving the lowest Mean Absolute Percentage Error of 0.63% on test data.
- The research addresses a critical industry need for the Electricity Generating Authority of Thailand, offering a decision support framework for high-voltage substation construction that satisfies both geotechnical stability requirements and electrical safety standards.
- We propose an innovative approach to hyperparameter optimization using Particle Swarm Optimization, systematically improving model architecture and training parameters.
- We developed an open-source web-based application enabling non-expert users to utilize deep learning capabilities. The model weights have been made publicly available to facilitate implementation or transfer learning in future research endeavors.

The remainder of this paper is structured as follows: Section 2 provides a comprehensive review of related works in tabular data modeling approaches, soil resistivity prediction, and attention-based architectures. Section 3 presents our methodology, detailing the soil testing procedures, dataset characteristics, and statistical properties of geotechnical parameters. Section 4 describes our dual-attention transformer architecture, including the feature embedding module, attention mechanisms, and prediction components. Section 5 outlines our experimental setup, feature preprocessing techniques, and the Particle Swarm Optimization framework for hyperparameter tuning. Section 6 presents our results, comparative model performance, and attention visualization analysis. Section 7 demonstrates the explainable aspects of our model through SHAP value interpretation and introduces our web-based application for practical engineering implementation. Finally, Section 8 concludes with a summary of contributions and directions for future research in explainable AI for geotechnical applications. Through this structure, we establish a robust foundation for understanding both the theoretical innovations and practical applications of our approach to soil resistivity prediction for high-voltage substation construction.

## 2. Relates works

Tabular data structures, characterized by their row-column organization, represent a fundamental data paradigm across multiple disciplines including biomedicine, economics, marketing, healthcare, and Internet of Things (IoT) applications (Chen and Guestrin, 2016). Accurate value prediction within these structures is essential for numerous applications including risk assessment, pharmaceutical discovery, customer segmentation analysis, clinical diagnostics, and network security protocols. Traditional approaches employed multilayer perceptrons (MLPs) for tabular data prediction (Alzo'ubi and Ibrahim, 2018; Youwai and Wongsala, 2024); however, these models demonstrated significant sensitivity to hyperparameter configurations and architectural specifications. Gradient-boosted decision trees (GBDTs) have consistently demonstrated superior performance in tabular data modeling, with implementations such as XGBoost (Chen and Guestrin, 2016; ForouzeshNejad et al., 2024; Tang, 2024; Zhou et al., 2024), CatBoost (Chehreh Chelgani et al., 2024; Prokhorenkova et al., 2018), and LightGBM (Bian et al., 2023; Ke et al., 2017; Sinha et al., 2023; Truong et al., 2024) employing sequential tree construction methodologies to optimize loss function minimization. These ensemble techniques exhibit exceptional predictive accuracy, computational efficiency, robust handling of heterogeneous data types, and scalability across large datasets. Gradient-boosted decision trees (GBDTs) always outperform the MLP architecture without attempt to optimize the hyperparameters.

The TabTransformer architecture (Huang et al., 2020) represents a significant methodological advancement in neural network applications for tabular data. This approach implements self-attention mechanisms to transform categorical feature embeddings into contextual representations, thereby enhancing predictive capabilities. Experimental validation across fifteen public datasets demonstrates substantial improvements in classification accuracy metrics compared to alternative neural architectures, performance equivalence with tree-based ensemble models, enhanced robustness to missing and noisy data points, superior interpretability characteristics, and marked predictive performance improvement in semi-supervised contexts utilizing unsupervised pre-training protocols. Subsequent research by Vyas (2024) has extended this framework through self-supervised learning paradigms that employ self-attention mechanisms to capture inter-feature dependencies, utilize surrogate supervised tasks for unlabeled data utilization, and demonstrate competitive performance against both traditional and contemporary methodological approaches. GatedTabTransformer (Cholakov and Kolev, 2022) enhances transformers with linear projections in MLP blocks and diverse activation functions, yielding significant performance gains in binary classification tasks with detailed optimization guidelines. The IoT Traffic Classification Transformer adapts this architecture for network security, using pre-training on MQTT datasets to achieve superior accuracy with minimal labeled data (Bazaluk et al., 2024).

Recent advances in foundation models for tabular data have yielded significant innovations, particularly the Tabular Prior-data Fitted Network (TabPFN), which employs a transformer-based architecture pre-trained on synthetic data generated via structural causal models. Hollmann et al. (2025) introduced TabPFN, a transformer-based foundation model for tabular data that outperforms traditional methods like gradient-boosted decision trees on datasets with up to 10,000 samples and 500 features. TabPFN leverages in-context learning (ICL) to autonomously learn effective strategies from synthetic tabular datasets, using a novel two-way attention mechanism optimized for the 2D nature of tables. Unlike traditional methods that require hand-engineered solutions for challenges like missing values and categorical data, TabPFN autonomously develops solutions by solving diverse synthetic tasks generated from structural causal models. The model significantly outperforms state-of-the-art baselines including CatBoost and XGBoost, even when these are tuned for hours, while requiring just seconds to make predictions. Beyond its predictive capabilities, TabPFN demonstrates foundation model qualities including fine-tuning, data generation, density estimation, and learning reusable embeddings, offering potential applications across domains from biomedicine to materials science.

The current investigation was motivated by the architectural framework of TabPFN, which implements one-dimensional dual attention mechanisms across both feature and batch dimensions in a bidirectional configuration. This approach was designed to elucidate patterns within and among features across the training data distribution. Our methodological approach extends this paradigm through the implementation of a two-dimensional attention mechanism and incorporation of an embedding layer to expand the dimensionality of individual features prior to attention processing. This architectural modification exhibits conceptual parallels to large language models, which utilize embedding transformations for token representation enhancement. We hypothesized that this structural modification would yield statistically significant improvements in model performance metrics. A critical component of our research methodology involves hyperparameter optimization, which constitutes a fundamental determinant of neural network efficacy. To address this challenge systematically, we employed particle swarm optimization (PSO) (Kennedy and Eberhart, 1995) to identify optimal architectural configurations and training hyperparameters, thereby maximizing predictive performance while maintaining computational efficiency constraints.

## 2. Model architecture

The model architecture is derived from TabPFN (Huang et al., 2020), as illustrated in Figs. 1 and 2. Initially, each input feature undergoes projection into a high-dimensional embedded vector space to capture complex feature relationships. This embedding transformation is essential for encoding categorical variables and enhancing representation capacity for continuous features, thereby facilitating the model's ability to detect non-linear patterns. Subsequently, a dual-attention mechanism is applied, consisting of feature attention and batch attention components operating in orthogonal dimensions (Fig.1). The feature attention module computes self-attention across feature embeddings, enabling the model to identify feature interactions and relative importance within each sample. This approach is particularly advantageous for tabular data where feature interdependencies are often complex and context-dependent. Concurrently, the batch attention module operates across samples, allowing the model to leverage inter-sample relationships and identify distributional patterns. This batch-wise attention effectively serves as an adaptive, data-driven regularization mechanism that mitigates overfitting by incorporating global dataset statistics into local predictions.

Following attention computation, dimensionality reduction is performed on both tensor streams through mean pooling operations, which aggregate the high-dimensional representations while preserving essential information (Fig. 2). Mean pooling was selected over alternative reduction methods (e.g., max pooling) due to its robustness to outliers and ability to maintain gradient flow during backpropagation, thereby stabilizing training dynamics. The reduced tensors are then concatenated along the feature dimension to form a unified representation that incorporates both feature-wise and batch-wise contextual information. This concatenation strategy enables the model to simultaneously leverage both attention mechanisms without imposing a hierarchical structure, thus preserving information from both pathways. The concatenated tensor is processed through a multilayer perceptron (MLP) with non-linear activation functions, systematically reducing the dimensionality to produce a scalar output value. The MLP serves as the final transformation that maps the attention-enriched representations to the regression target, with its depth providing sufficient capacity to approximate complex functions while maintaining computational efficiency.

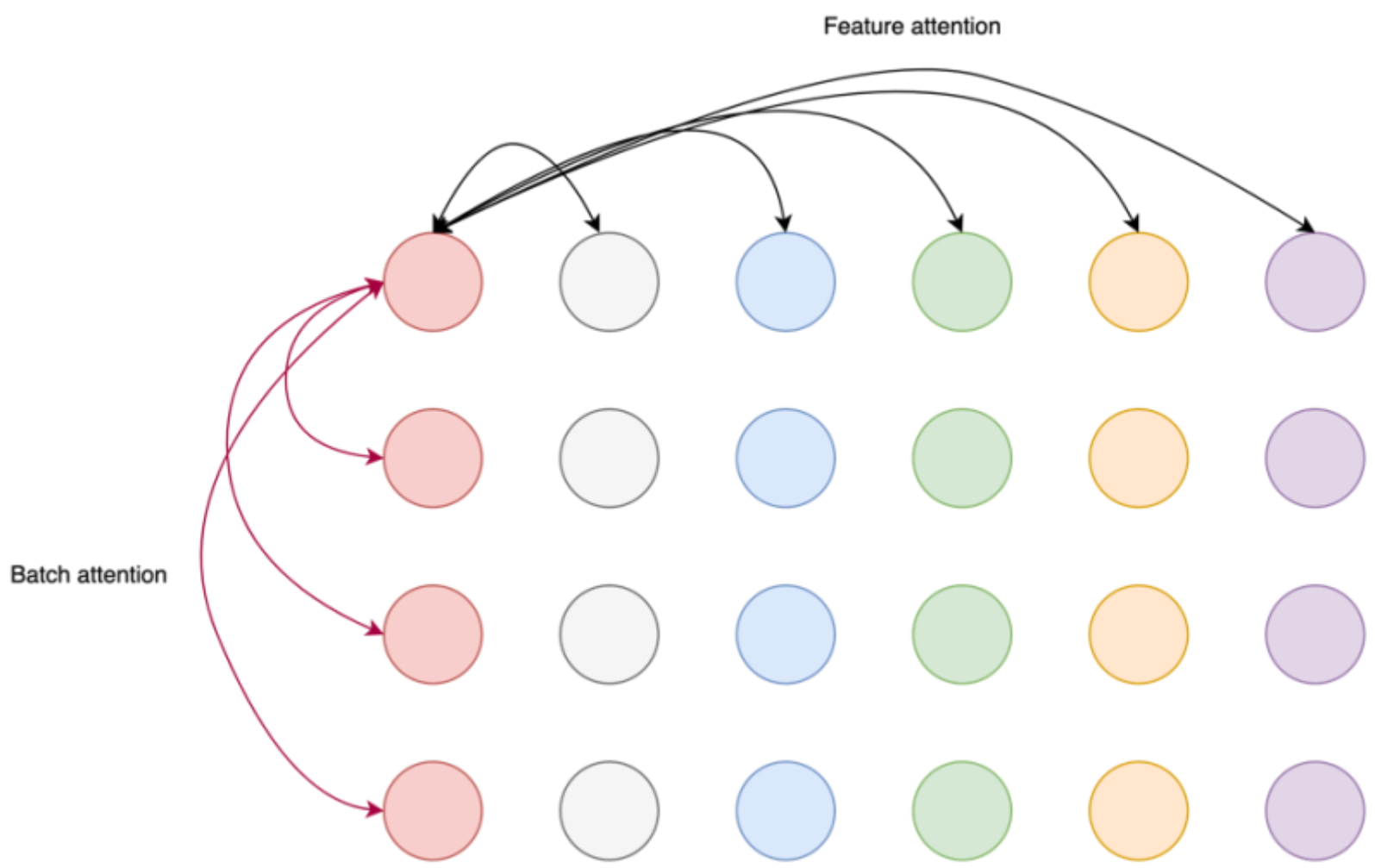

**Fig. 1** The dual attention of the proposed model

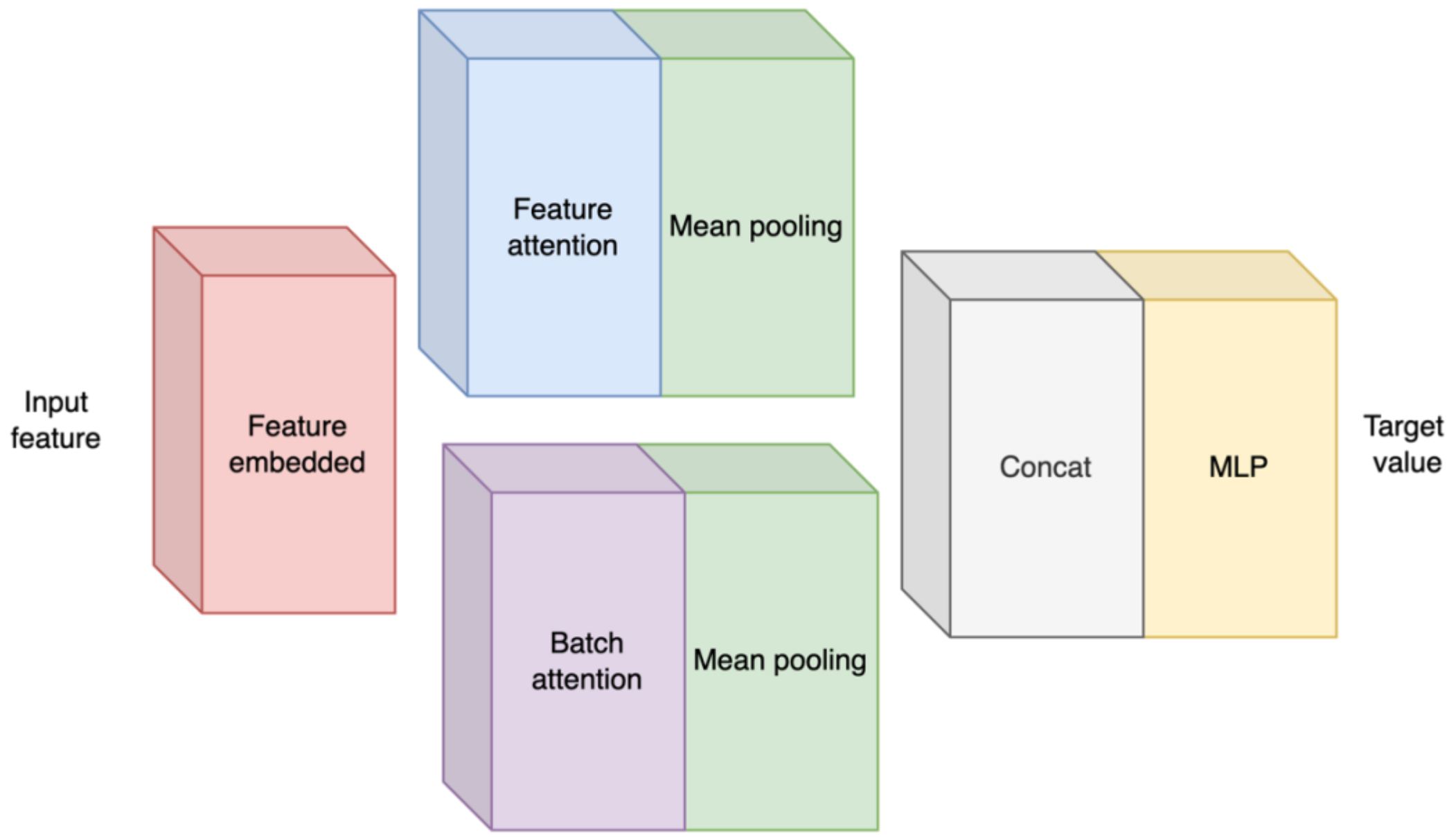

**Fig. 2** The model architecture

The model consists of three primary components:
1. Feature Embedding Module: Projects individual input features into a higher-dimensional space.
2. Dual Attention Mechanism: Applies multi-head self-attention across features and samples.
3. Prediction Module: Combines attention outputs and generates final predictions via an MLP.

The model's parameters include:
-$N_f$: Number of input features (default: 6).
-$D_e$: Embedding dimension (default: 36).
-$D_o$: Output dimension (default: 1).
-$H$ Number of attention heads (default: 4).

## 2.1 Feature Embedding Module

For an input tensor($X \in \mathbb{R}^{\mathrm{B} \times \mathrm{N_f}}$), where B is the batch size and $N_f$ is the number of features, each feature $x_{:,i} \in \mathbb{R}^B$ for( $i = 0, 1, \ldots, N_f - 1$) processes independently by a 'FeatureEmbedding` module. The embedding function for the $i - th$ feature is defined as Equation 1:

$$E_i(x_{:,i}) = W_{i,2} \cdot \mathrm{ReLU}(W_{i,1} \cdot x_{:,i} + b_{i,1}) + b_{i,2} \tag{1}$$

where:
- $W_{i,1} \in \mathbb{R}^{D_e \times \mathbb{1}}, b_{i,1} \in \mathbb{R}^{D_e}$: Parameters of the first linear layer.

- $W_{i,2} \in \mathbb{R}^{D_e \times D_e}, b_{i,2} \in \mathbb{R}^{D_e}$: Parameters of the second linear layer.
$- E_i(x_{:,i}) \in \mathbb{R}^{B \times D_e}$ : Embedded representation of the i $-$ th feature.

The embedded features are stacked to form $E = \left[E_0, E_1, \dots, E_{\{N_f-1\}}\right] \in \mathbb{R}^{\{N_f \times B \times D_e\}}$.

## 2.2 Dual Attention Mechanism

2.2.1 Feature Attention
Feature attention enables each feature to attend to all other features within a sample. The multi-head self-attention mechanism is applied over the feature dimension of $E$ For $L_f = 4$ iterations, the process is (Equations 2-4):

$$F^{(0)} = E \quad (2)$$
$$F^{(l+1)} = F^{(l)} + \text{MultiHeadAttention}\left(F^{(l)}, F^{(l)}, F^{(l)}\right) \quad (3)$$
$$F^{(l+1)} = \text{LayerNorm}\left(F^{(l+1)}\right) \quad (4)$$

where:
- $\text{MultiHeadAttention}(Q, K, V)$ computes attention with $H$ heads, as defined in (Vaswani et al., 2017):

$$\text{Attention}(Q, K, V) = \text{softmax}\left(\frac{QK^T}{\sqrt{D_e/H}}\right)V \quad (5)$$

-$F^l \in R^{N_f \times B \times D_e}$: Feature-attended output at iteration $l$.
- Final output: $F = F^{(L_f)}$

The residual connection and layer normalization stabilize training and enhance gradient flow.

## 2.3 Sample Attention

Sample attention operates across the batch dimension, allowing samples to attend to one another. The input tensor $E$ is permuted to $S_{\text{in}} = E^T \in R^{B \times N_f \times D_e}$ then re-permuted to $S'_{\text{in}} = \text{permute}(S_{\text{in}}, [1,0,2])$ in $\mathrm{R}^{\mathrm{N_f \times B \times D_e}}$ . For $(L_s = 4)$ iterations (Equations 6-7):

$$S^{(0)} = S'_{\text{in}} \quad (6)$$
$$S^{(l+1)} = S^{(l)} + \text{MultiHeadAttention}\left(S^{(l)}, S^{(l)}, S^{(l)}\right) \quad (7)$$
$$S^{(l+1)} = \text{LayerNorm}\left(S^{(l+1)}\right) \quad (8)$$

where:
-$S^{(l)} \in R^{N_f \times B \times D_e}$: Sample-attended output at iteration $l$
- Final output: $S = S^{(L_s)}$

## 2.4 Combination and Prediction

The outputs of feature and sample attention are permuted back to $F' = F^T \in \mathbb{R}^{B \times N_f \times D_e}$ and $S' = S^T \in \mathbb{R}^{B \times N_f \times D_e}$. Mean pooling is applied across the feature dimension (Equations 9-10):

$$F_{\text{pooled}} = \text{mean}(F', \text{dim} = 1) \in R^{B \times D_e} \quad (9)$$
$$S_{\text{pooled}} = \text{mean}(S', \text{dim} = 1) \in R^{B \times D_e} \quad (10)$$

The pooled representations are concatenated and projected (Equations 11):

$$C = \text{ReLU}\left(W_c \cdot \left[F_{\text{pooled}}, S_{\text{pooled}}\right] + b_c\right) \quad (11)$$

where:
- $\left[F_{\text{pooled}}, S_{\text{pooled}}\right] \in \mathbb{R}^{B \times 2D_e}$ : Concatenated tensor.

- $W_c \in \mathbb{R}^{D_e \times 2D_e}, b_c \in \mathbb{R}^{D_e}$: Combination layer parameters.
- $C \in R^{B \times D_e}$: Combined representation.

The final prediction is computed via an MLP (Equation 12):

$$Y = W_4 \cdot \text{ReLU}(W_3 \cdot \text{ReLU}(W_2 \cdot \text{ReLU}(W_1 \cdot C + b_1) + b_2) + b_3) + b_4 \quad (12)$$

where:
- $W_1 \in \mathbb{R}^{128 \times D_e}, b_1 \in \mathbb{R}^{128}$
- $W_2 \in \mathbb{R}^{64 \times 128}, b_2 \in \mathbb{R}^{64}$
- $W_3 \in \mathbb{R}^{32 \times 64}, b_3 \in \mathbb{R}^{32}$
$-W_4 \in \mathbb{R}^{D_o \times 32}, b_4 \in R\mathbb{R}^{D_o}$
-$Y \in \mathbb{R}^{B \times D_o}$: Final output.

The proposed model leverages the strengths of transformers by applying attention in two complementary directions: across features and across samples. This dual mechanism captures both local feature interactions and global sample relationships, making it particularly suited for tabular datasets with complex dependencies. The iterative attention loops ($L_f = L_s = 4$) enhance the model's capacity to learn hierarchical representations.

## 3. Data characteristics

### 3.1 Laboratory Testing

In this investigation, seven distinct soil types from various regions across Thailand were examined: (1) KMUTT sand (cleaned sand passed through sieve No. 40 and retained on sieve No. 100), (2) Loei soil, (3) Bang Pakong soil, (4) Roi Et soil, (5) Ubon Ratchathani soil, (6) Chaiyaphum soil, and (7) Nakhon Ratchasima soil, as illustrated in Fig 3. KMUTT sand is extensively utilized in geotechnical engineering experiments conducted at the Geotechnical Engineering Laboratory, King Mongkut's University of Technology Thonburi (KMUTT) (e.g., (Chantachot et al., 2016)(Dararat et al., 2021); (Jariyatatsakorn et al., 2024); (Kongkitkul et al., 2011)). According to geotechnical classification, the soils fall into three categories: poorly graded sand (SP), poorly graded silty sand (SP-SM), and poorly graded clayey sand (SP-SC). The geographical distribution of the sampling sites is shown in Fig. 3, spanning from the central region (KMUTT sand near Bangkok) to the northern region (Loei soil) and throughout the northeastern provinces (Bang Pakong, Roi Et, Ubon Ratchathani, Chaiyaphum, and Nakhon Ratchasima soils). This strategic sampling approach enables comprehensive characterization of Thailand's heterogeneous soil profiles.

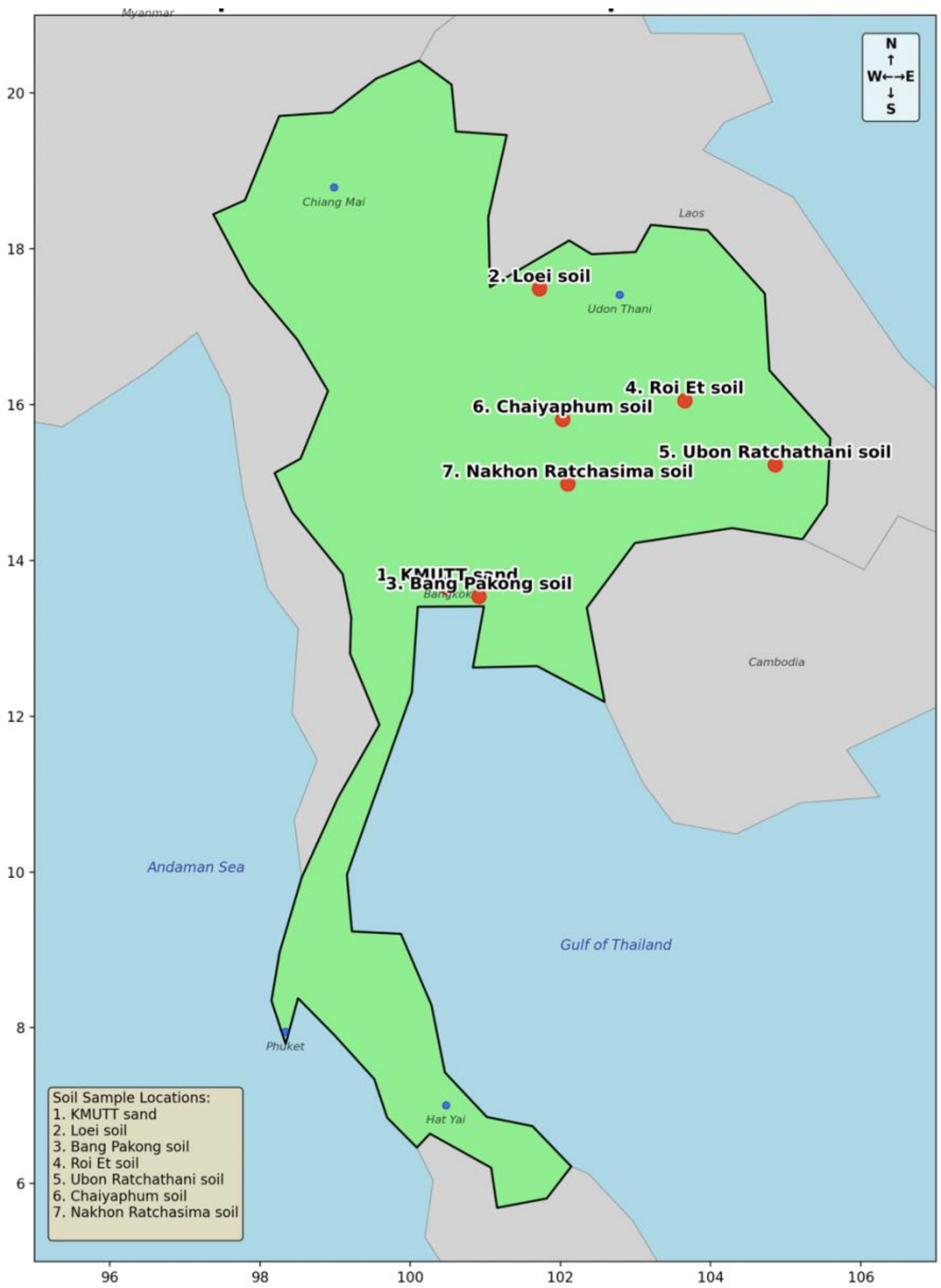

**Fig. 3** The location of soil in this study

The electrical resistivity of the seven Thai soil samples is governed by their distinct physicochemical properties presented in Table 1. Particle size distribution parameters, particularly mean diameter ($D_{50}$ :0.189-0.754 mm) and uniformity coefficients (Cu: 1.335-3.931), influence pore connectivity and resistivity pathways. Fines content (F200: 0-9.80%) represents a parameter that varies across the samples, including KMUTT sand (0% fines) at one end of the spectrum. Plasticity indices ranging from non-plastic to 14.53% indicate variable clay mineral content, with Bang Pakong, Chaiyaphum, and Nakorn Ratchasima soils exhibiting different Plasticity index (PI) values that affect surface conductivity mechanisms. The soil classification differences between SP-SC and SP-SM types provide additional variables for examining electrical behavior, with the contrasting properties of clay versus silt particles offering different conduction mechanisms. These relationships enable comprehensive investigation across the parameter space for geotechnical applications including electrical resistivity tomography surveys, grounding system design, and corrosion potential assessments in Thai geotechnical engineering contexts.

**Table 1** The soil properties in this study

| No. | Materials | Gs | PI (%) | $F_{200}$ (%) | Soil type |
|---|---|---|---|---|---|
| 1 | KMUTT sand | 2.665 | - | 0 | SP |
| 2 | Loei soil | 2.755 | 1.01 | 9.80 | SP-SC |
| 3 | Bang Pakong soil | 2.728 | 14.53 | 6.14 | SP-SM |
| 4 | Roi Ed soil | 2.634 | 5.58 | 7.96 | SP-SC |
| 5 | Ubon Ratchathani soil | 2.653 | 0.46 | 9.06 | SP-SC |
| 6 | Chaiyaphum soil | 2.657 | 13.09 | 9.71 | SP-SM |
| 7 | Nakorn Ratchasima soil | 2.645 | 13.85 | 9.09 | SP-SM |

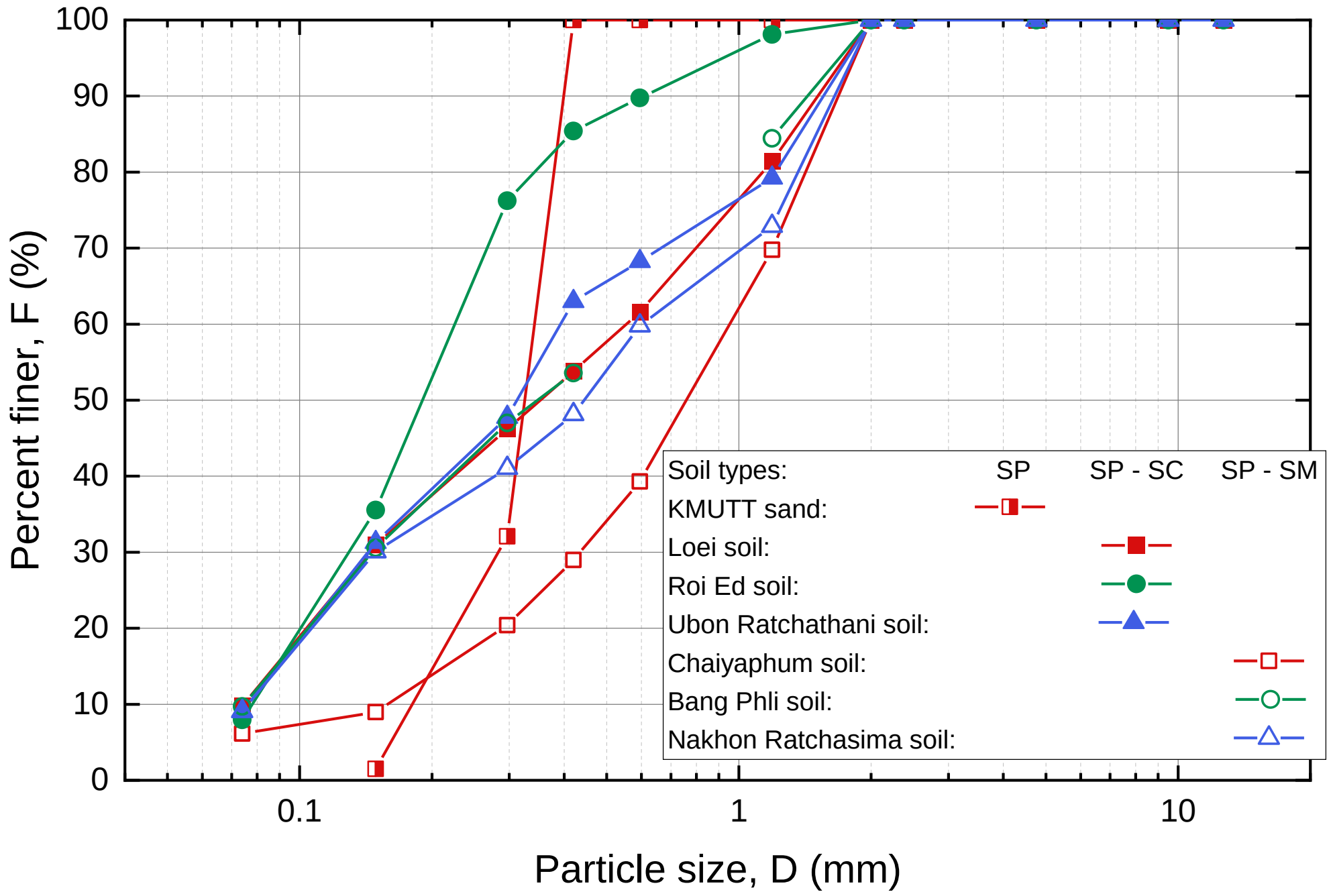

**Fig. 4** Grain size distribution of soil

The presented particle size distribution curves for various soil types (Fig. 4), plotted as percent finer (F) versus particle diameter (D) on a semi-logarithmic scale, provide critical data for geotechnical characterization. Besides KMUTT sand, whose particle distribution was controlled to be uniform, the other soil types were naturally obtained and exhibit distinctive gradation patterns. Roi Et soil shows the steepest curve, with 35% finer at 0.2 mm and 98% at 2 mm, indicating a relatively uniform particle distribution. In contrast, Chaiyaphum soil displays a more gradual slope, suggesting a wider particle size distribution. These soils, classified as SP (poorly graded sand), SP-SC (poorly graded sand with clay), and SP-SM (poorly graded sand with silt) according to USCS, exhibit varying fines content, which directly influences electrical conductivity pathways. The correlation between grain size distribution and electrical resistivity is fundamental since the proportion of fine particles relates to lower resistivity values due to enhanced surface conductivity mechanisms, while effective porosity, governed by particle distribution, determines pore fluid volume and connectivity—primary factors in soil electrical conduction. Consequently, Roi Ed soil, with its higher percentage of fine particles, would likely exhibit lower electrical resistivity than coarser materials like KMUTT sand, especially when saturated, making these granulometric analyses essential components for developing empirical correlations between soil physical properties and electrical characteristics for accurate geophysical subsurface modeling.

Heterogeneous soil samples were selected to quantify electrical resistivity variations as a function of soil composition and granulometric distribution. Electrical resistivity measurements were conducted using a four-electrode testing apparatus with constant current input to characterize resistance properties under controlled laboratory conditions (Fig. 5) according to ASTM G57 (ASTM International, 2020). Soil volumetric water content was systematically manipulated as an independent variable to assess its influence on electrical resistivity, a critical parameter for evaluating grounding system performance in electrical transmission infrastructure. KMUTT sand was utilized as a control specimen due to its homogeneous particle size distribution and minimal ionic concentration, while the remaining soil samples represented typical field conditions at various EGAT transmission tower installations throughout Thailand. Electrical resistivity tests were performed at standardized compaction levels to approximate in-situ density conditions, facilitating the development of empirical correlations between laboratory measurements and field applications for electrical grounding system design.

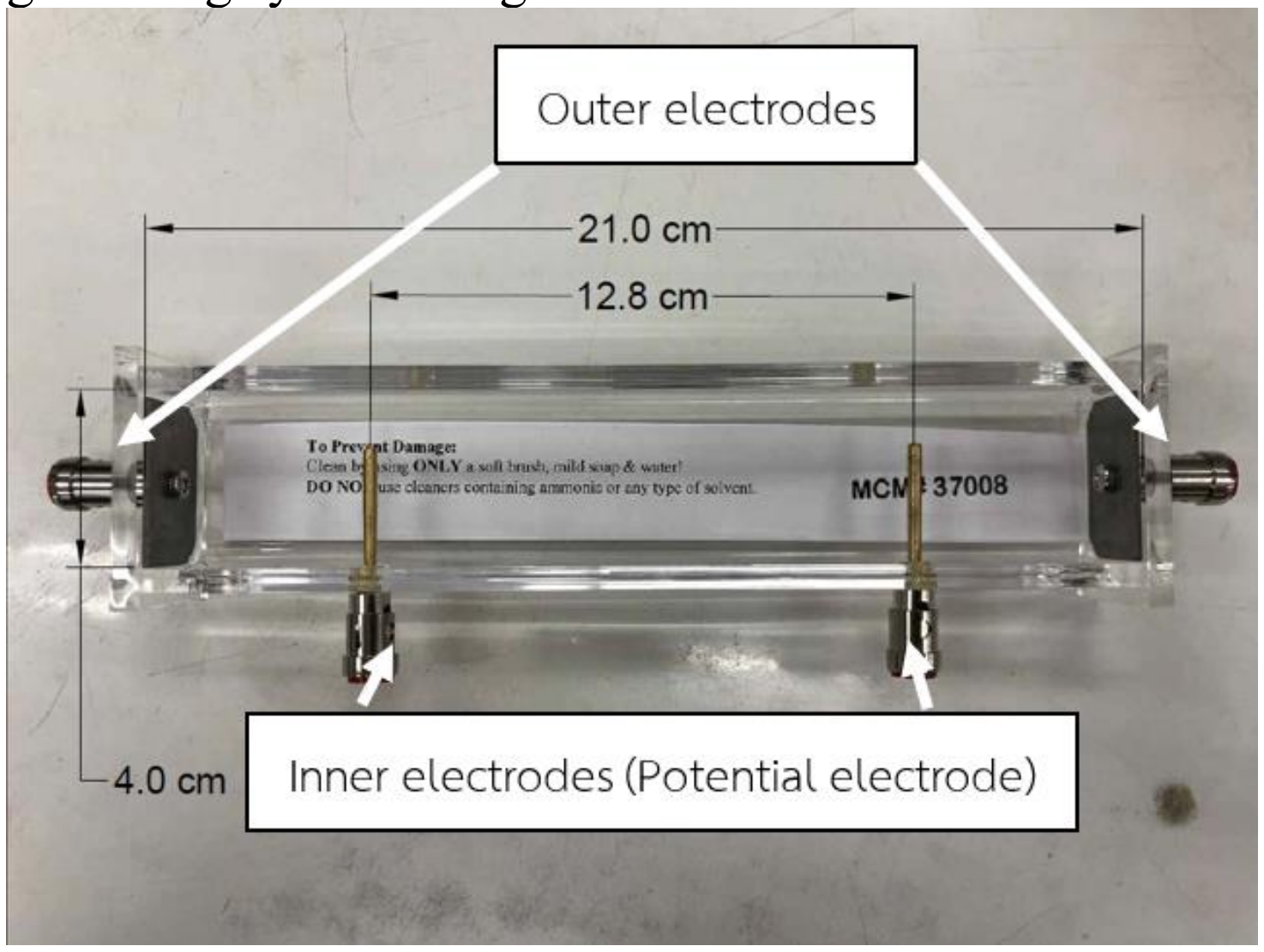

**Fig. 5** Soil box for electrical resistivity test

For each of the seven soil types investigated, six discrete water content values were established. Specimens at each water content value were subsequently compacted to achieve six distinct dry density targets, resulting in a 6×6 experimental matrix per soil type. The selected water content values were derived from the arithmetic mean of the optimum moisture content determined from standard and modified Proctor compaction tests. Maximum dry density values from these tests were not utilized as control parameters due to sample preparation constraints. This systematic variation of both moisture content and dry density was implemented to comprehensively investigate the electrical resistivity behavior of soils across a wide range of physical states. The extensive experimental matrix was necessary because soil electrical resistivity is known to be highly sensitive to both moisture content and compaction state, with previous research demonstrating non-linear relationships between these parameters. Additionally, the combined influence of varying both water content and dry density simultaneously provides critical insights into the complex electro-physical soil behavior that cannot be adequately characterized through single-variable experimentation, particularly for geotechnical engineering applications requiring precise resistivity measurements. This experimental design yielded 36 unique conditions (6 dry densities × 6 water contents) per soil type, totaling 252 experimental cases.

Soil resistivity tests were conducted on various specimens, differentiated by dry density, water content, and soil type. Direct current (DC) electricity was applied to the two outer electrodes (Fig. 5), while the DC current (I) and DC voltage (V) were measured at the two inner electrodes. In this study, the applied DC voltage at the outer electrodes varied across six levels: 5, 10, 15, 20, 25, and 30 volts, resulting in a total of 1,512 trials (252 × 6). Each measured pair of I and V values was used to calculate electrical resistance (R). The calculated R values from different I-V pairs showed only slight variations, so the average R was determined. The soil resistivity (ρ) was then calculated from this averaged R and is employed for data analysis.

## 3.2 Statistical data properties

The soil properties shown in the distributions significantly influence electrical resistivity measurements through their combined effects (Fig. 6). The relatively consistent dry density (ρd) of 1.52 g/cm$^3$ provides a baseline compaction level, while the moderate moisture content (w) averaging 13.23% directly enhances electrical conductivity by facilitating ion movement through the soil matrix. This conductivity is further affected by the degree of saturation (S) at 47.99%, though its high variability (std: 15.52%) suggests zones of differing resistivity throughout the soil mass because electrons follow paths of least resistance through water-filled pores. The void ratio (e) averaging 0.78 controls how water and dissolved ions navigate through the soil structure, as smaller and more tortuous pathways increase resistivity while more connected pore spaces decrease it. This works in conjunction with the consistent mineral composition indicated by the specific gravity (Gs) of 2.68, which suggests the presence of common soil minerals like quartz and feldspar that typically exhibit high resistivity unless altered by conductive coatings or weathering. Additionally, the presence of fine particles (F200 averaging 7.39%) and clay content suggested by the plasticity indices (LL: 24.99%, PL: 18.06%) further enhances conductivity through greater surface area, ion exchange capacity, and the formation of electrical double layers at clay particle surfaces that facilitate charge movement. The bimodal distribution in the liquid limit (LL) graph indicates two distinct soil types or layers might be present, potentially creating resistivity contrasts detectable during testing. Together, these interrelated properties create a soil with moderate to good electrical conductivity characteristics, though spatial variations in resistivity should be expected due to the heterogeneity in moisture distribution, saturation levels, and the apparent presence of different soil types as suggested by the multimodal distributions in several parameters.

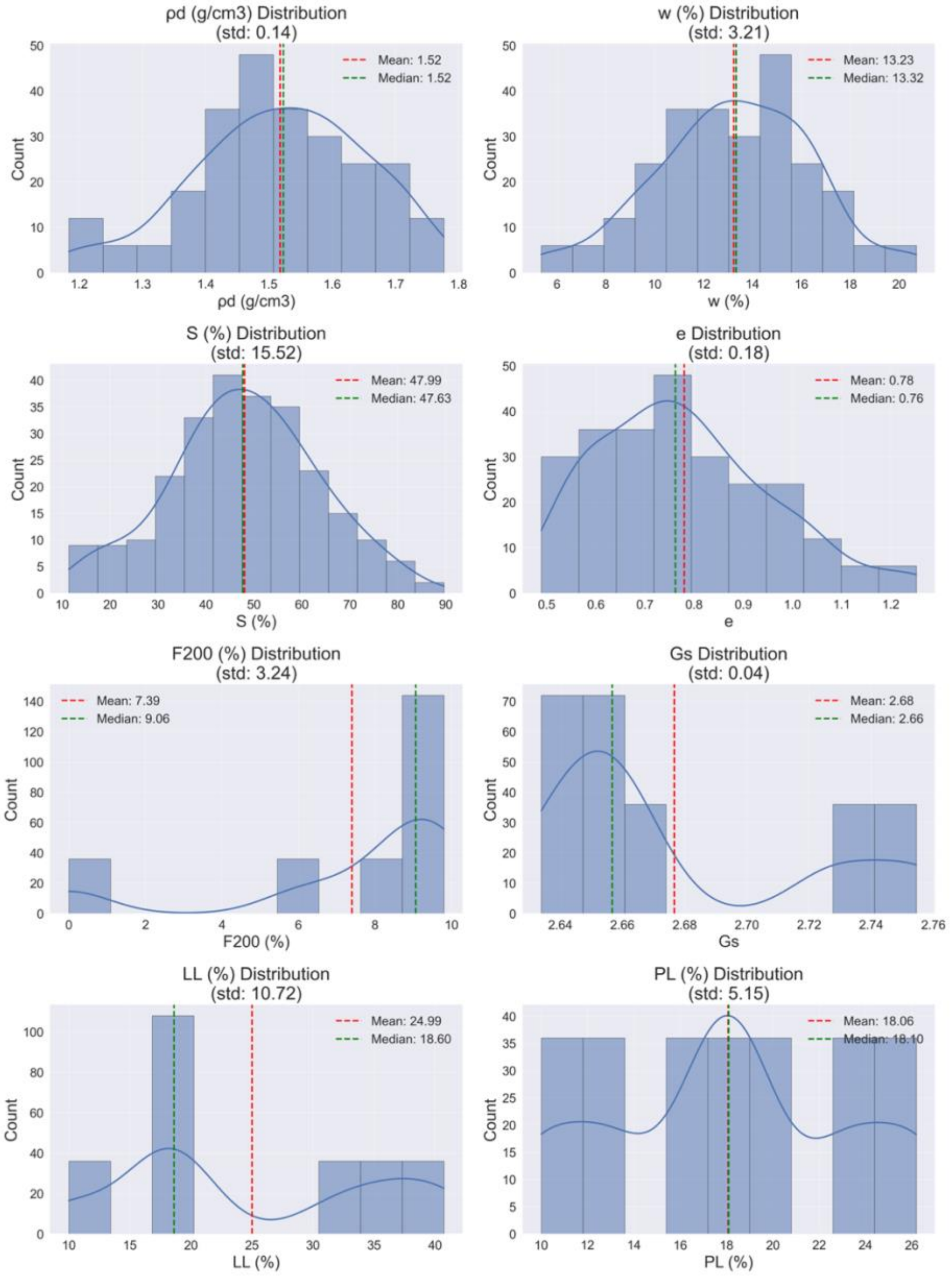

**Fig. 6** Statistical data properties

The Spearman correlation heatmap reveals critical relationships between soil properties and electrical resistivity (Y) as shown in Fig. 7. The negative correlation between resistivity and moisture-related parameters (w: -0.19, S: -0.31) confirms that as soil moisture and saturation increase, electrical resistivity decreases due to water's role as a primary conductor of electricity through the soil matrix. The positive correlation with void ratio (e: 0.17) supports this relationship, as higher void ratios typically mean less compaction and potentially drier conditions, leading to increased resistivity. The weak positive correlation with fine content (F200: 0.11) might seem counterintuitive, as fine particles typically enhance conductivity. However, this could indicate that in this specific soil dataset, the effect of particle size distribution is overshadowed by moisture content variables or that the fines present are less conductive minerals. The negative correlations with plasticity indices (LL: -0.17, PL: -0.08) suggest that clay content does contribute to decreased resistivity, likely through surface conductivity mechanisms and enhanced ion exchange capacity. The correlation between specific gravity (Gs) and resistivity (-0.15) further supports the influence of mineral composition, with denser minerals potentially containing more conductive elements. The strongest negative correlation with saturation degree (S: -0.31) emphasizes that the continuity of water films in soil pores is the dominant factor controlling electrical pathways through the soil. The moderate strength of these correlations (none exceeding ±0.31) indicates that soil electrical resistivity is influenced by a complex interplay of multiple factors rather than being dominated by any single property. This reinforces the importance of considering the full suite of soil properties when interpreting resistivity measurements for geotechnical applications, as the relationships are neither simple nor linear but represent a multifaceted system where moisture, compaction, and composition work together to determine the soil's electrical characteristics.

The feature set employed in the established model was subsequently reduced to 6 parameters, as two parameters—void ratio and degree of saturation—can be directly calculated from specific gravity (Gs), water content, and dry density through established geotechnical relationships. This reduction is methodologically justified as these derived parameters represent redundant information within the model's feature space. Furthermore, these two parameters present practical implementation challenges, as they require additional computational steps that may be inconvenient for practicing engineers when compared to the more directly measurable soil phase variables (density, specific gravity, and water content). The decision to exclude these derived parameters enhances model parsimony while maintaining the physical interpretability of the input feature set, facilitating more straightforward application in geotechnical engineering practice.

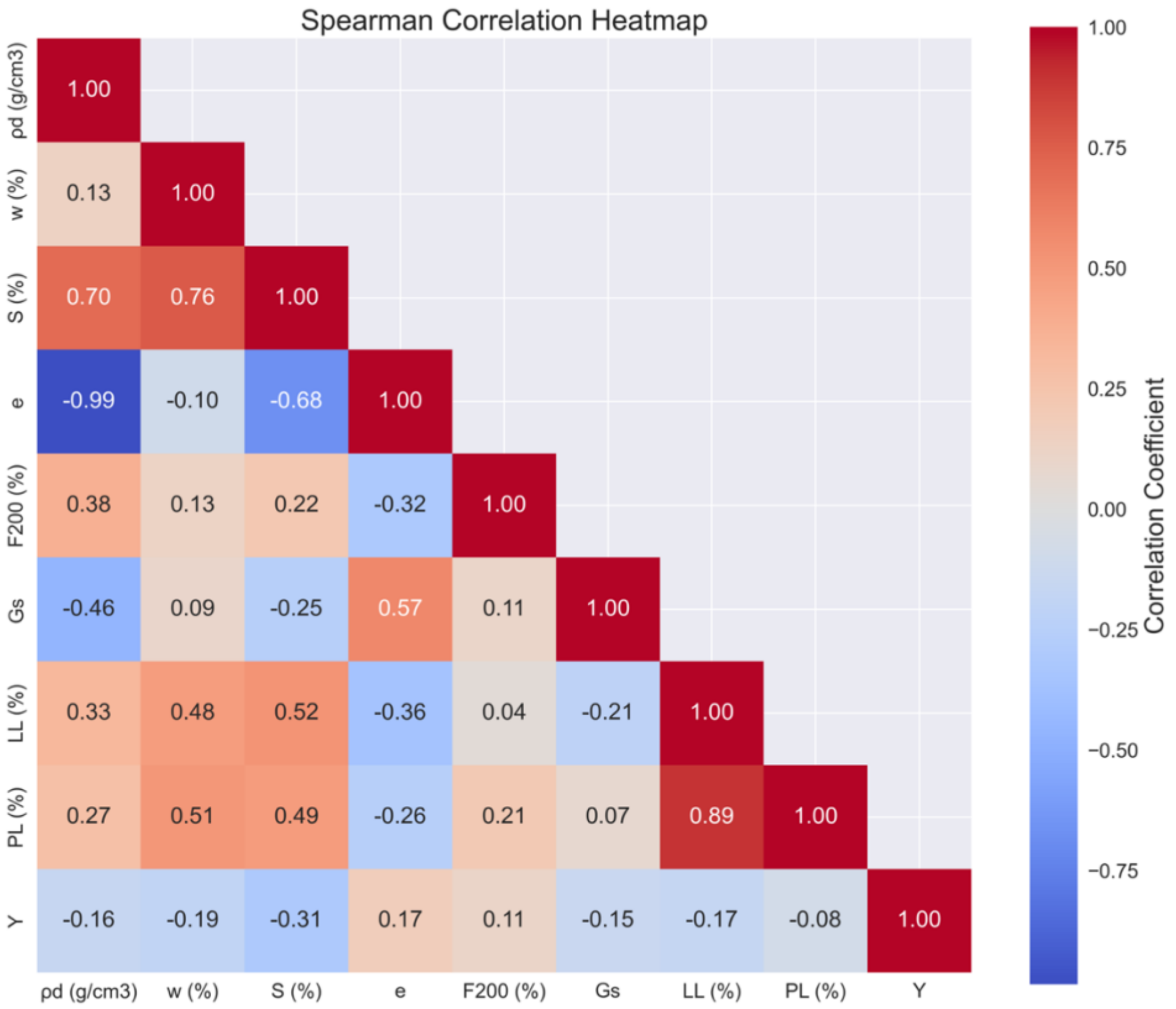

**Fig.7** Correlation heat map of feature to soil resistivity

## 4. Experiment

In this study, we use Yeo-Johnson power transformation (Riani et al., 2023) to preprocess our feature data prior to model training. This transformation was selected due to its robust capability to normalize skewed distributions while accommodating both positive and negative values in our dataset. The Yeo-Johnson transformation is mathematically defined as (Equation 13):

$$\psi(x,\lambda) = \begin{cases} frac(x+1)^{\lambda} - 1 & \text{if } x \geq 0, \lambda \neq 0 \\ \ln(x+1) & \text{if } x \geq 0, \lambda = 0 \\ -\frac{[(-x+1)^{2-\lambda}-1]}{2-\lambda} & \text{if } x < 0, \lambda \neq 2 \\ -\ln(-x+1) & \text{if } x < 0, \lambda = 2 \end{cases} \quad (13)$$

Our implementation employs the scikit-learn PowerTransformer(method='yeo-johnson') which optimizes the λ parameter for each feature independently by maximizing the log-likelihood function (Equation 14):

$$L(\lambda) = -\frac{n}{2}\ln\left(\frac{1}{n}\sum_{i=1}^{n}(y_i - \bar{y})^2\right) + (\lambda - 1)\sum_{i=1}^{nsgn}(x_i)\ln(|x_i| + 1) \quad (14)$$

The primary rationale for employing this transformation stems from the non-Gaussian nature of our input features, which exhibited significant skewness and heteroscedasticity in preliminary data analysis. By applying Yeo-Johnson transformation, we address several critical requirements of our statistical modeling approach: (1) normalization of feature distributions to satisfy the Gaussian assumption underlying our parametric models, (2) stabilization of variance across the feature domain to ensure reliable confidence intervals in our predictions, (3) mitigation of outlier influence without data removal, and (4) potential linearization of complex relationships between predictors and response variables. Furthermore, the transformation's ability to handle mixed-sign data without artificial shifting was particularly valuable for features that naturally span both positive and negative domains in our dataset.

The dataset (n=252) is partitioned into training, testing, and validation subsets using an 80/10/10 ratio distribution. The training and validation datasets were utilized to monitor the model's learning progression and prevent overfitting during the development phase, while the independent test dataset (10%) was reserved exclusively for evaluating the final model performance through established performance metrics. This strict separation methodology follows statistical best practices by preventing data leakage between development and evaluation phases, thereby producing more reliable generalization estimates. The relatively large training proportion (80%) ensures sufficient data for the model to learn complex patterns within the soil resistivity relationships, while the balanced allocation between validation and testing (10% each) provides adequate statistical power for both hyperparameter tuning and final performance assessment without compromising either process. The validation subset serves as a proxy for unseen data during training, enabling early stopping and model selection without contaminating the test set, thus maintaining the scientific integrity of the reported performance metrics by eliminating potential optimization bias that could occur if parameters were tuned directly on the test data. The model training procedure employs the Mean Squared Error (MSE) as the loss function( $\mathcal{L}$) (Equation 15):

$$\mathcal{L} = \frac{1}{n}\sum_{i=1}^{n}(y_i - \hat{y_i})^2 \quad (15)$$

where $y_i$ represents the actual resistivity value and $y_i$ represents the predicted value for the $i$-th sample. The Adam (Kingma and Ba, 2014) optimizer was selected for parameter updates. The training procedure was structured as follows:

The model's performance was comprehensively evaluated using two complementary metrics: Mean Absolute Percentage Error (MAPE) and the Coefficient of Determination ($R^2$) as shown in Equations 16 and 17. MAPE normalizes errors relative to actual values, offering a percentage-based score to identify systematic over- or underestimation biases and ensure scale-invariant interpretability, particularly in applications where proportional accuracy is prioritized. $R^2$ assesses the proportion of variance in the target variable explained by the model, with values near 1.0 indicating strong explanatory power and effective pattern capture. Together relative error impact (MAPE) and global trend alignment ($R^2$), enabling robust identification of both local prediction precision and global model fit. This multi-metric approach ensures alignment with application-specific requirements for reliability and generalizability, safeguarding against over- or underestimation of model performance (Equations 15 and 16).

$$\text{MAPE} = \frac{100\%}{n}\sum_{i=1}^{n}\left|\frac{y_i - \hat{y}_i}{y_i}\right| \tag{16}$$

$$R^2 = 1 - \frac{\sum_{i=1}^{n}\left(y_i - \hat{y}_i\right)^2}{\sum_{i=1}^{n}\left(y_i - \bar{y}\right)^2} \tag{17}$$

This study implements Particle Swarm Optimization (PSO) (Kennedy and Eberhart, 1995) for hyperparameter optimization of a tabular transformer neural network architecture. PSO represents an efficient meta-heuristic approach for navigating high-dimensional, non-convex search spaces without requiring gradient information. The algorithm's population-based structure enables parallel exploration of the hyperparameter space while maintaining computational efficiency. The canonical PSO algorithm maintains a population of candidate solutions (particles) that traverse the search space according to their individual velocities. Each particle i retains memory of its best historical position $p_i$ and is influenced by the swarm's global best position $g$. The position and velocity update mechanisms are governed by the following equations (Equations 18 and 19):

$$v_i(t+1) = wv_i(t) + c_1 r_1[p_i - x_i(t)] + c_2 r_2[g - x_i(t)] \tag{18}$$

$$x_i(t+1) = x_i(t) + v_i(t+1) \tag{19}$$

where:
$v_i(t)$ and $x_i(t)$ represent the velocity and position of particle i at iteration $t$
$w$ denotes the inertia weight (set to 0.5)
$c_1$ and $c_2$ are the cognitive and social coefficients respectively (both set to 1.5)
$r_1$ and $r_2$ are uniform random variables in the range [0,1]

The inertia weight $w$ controls the momentum of particles, while the cognitive coefficient $c_1$ governs the influence of the particle's memory, and the social coefficient $c_2$ determines the swarm's influence on individual particles. The implemented PSO framework optimizes the following hyperparameters within the specified bounds (Table 2):

**Table 2** The range of the hyperparameters tuning from Particle Swarm Optimization (PSO)

| Hyperparameter | Lower bound | Upper bound | Optimized value |
|---|---|---|---|
| **Embedding dimension** | 16 | 64 | 36 |
| **Attention heads** | 1 | 8 | 4 |
| **Attention loops** | 2 | 12 | 4 |
| **Hidden dimension 1** | 32 | 256 | 128 |
| **Hidden dimension 2** | 16 | 128 | 64 |
| **Hidden dimension 3** | 8 | 64 | 32 |
| **Batch size** | 8 | 32 | 55 |
| **Learning rate** | 0.00001 | 0.01 | 0.00086 |

The optimization procedure employs a swarm of 10 particles iterated over 15 generations, with each particle's fitness evaluated using the coefficient of determination ($R^2$) on a validation set. To ensure proper model construction, additional constraints are enforced, such as requiring the embedding dimension to be divisible by the number of attention heads. The fitness function is defined as the negative $R^2$ score to transform the maximization problem into a minimization one (Equation 20):

$$f(x_i) = -R^2(x_i) \tag{20}$$

where $R^2(\boldsymbol{x_i})$ is the coefficient of determination achieved by the model with hyperparameters represented by position $\boldsymbol{x_i}$. For each particle evaluation: (1) A tabular transformer model is instantiated with the particle's hyperparameter configuration; (2) The model undergoes training for 100 epochs using Adam optimization with the particle's specified learning rate; (3) The best $R^2$ score achieved during training is recorded as the particle's fitness; (4) Personal and global best positions are updated accordingly. The algorithm incorporates boundary handling techniques to ensure hyperparameters remain within the specified ranges. Additionally, mixed continuous-discrete parameter handling is implemented, with integer parameters like batch size and attention heads rounded to the nearest integer after position updates. The PSO optimization progress is monitored through multiple metrics including best $R^2$ score per iteration, validation loss of the best model per iteration, and convergence characteristics of the swarm. Visualization of the optimization trajectory provides insights into the algorithm's exploration-exploitation behavior and the hyperparameter landscape's characteristics. The final optimized model demonstrates superior performance compared to manually tuned configurations, validating the efficacy of the PSO approach for neural network hyperparameter optimization.

The training and validation loss trajectories depicted in Figure 8 demonstrate optimal convergence characteristics across 1000 epochs of model training. Initial loss values approaching 1.0 undergo rapid exponential decay, stabilizing at approximately 0.01 within the first 50 epochs. Notably, the convergence behavior exhibits remarkable stability throughout the subsequent training period, with minimal oscillatory patterns observed in both metrics. The model achieved optimal generalization performance at epoch 827, as indicated by the minimum validation loss value (highlighted). The consistent parallelism between training and validation curves suggests effective hyperparameter optimization, particularly regarding regularization mechanisms. This absence of divergence between the metrics provides strong evidence against overfitting phenomena. A transient perturbation in validation loss appears near epoch 800, but the system rapidly returns to equilibrium, further confirming the robustness of the selected hyperparameters. These results demonstrate that the optimization strategy successfully addressed the bias-variance tradeoff, producing a model with exceptional generalization capabilities. The observed convergence pattern aligns with theoretical expectations for properly regularized deep learning architectures when subjected to systematic hyperparameter tuning methodologies.

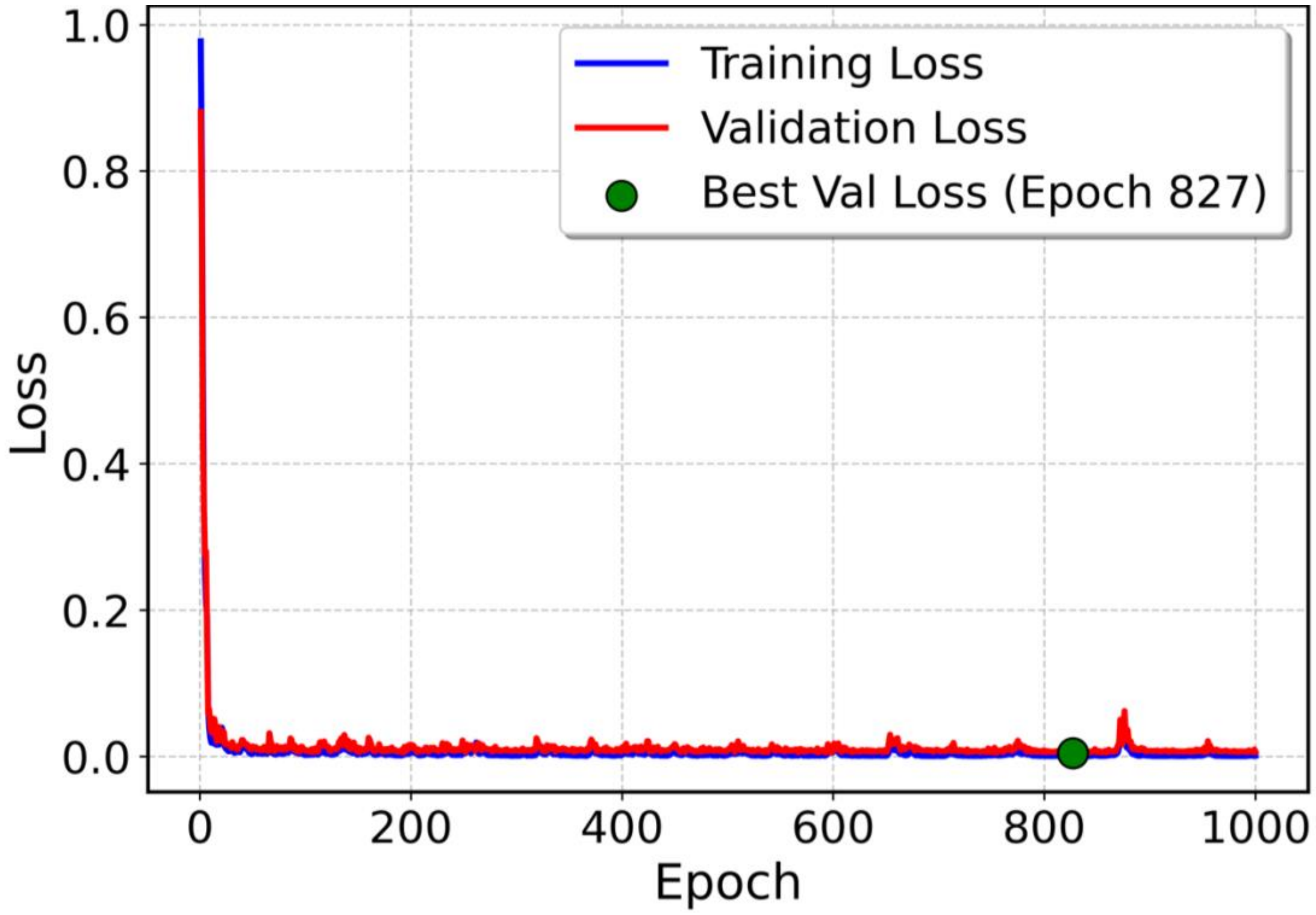

**Fig. 8** Training Loss vs. Validation Loss over Epochs

The Fig. 9 presents a scatter plot comparing predicted values against actual values across training (blue circles), validation (green triangles), and test (red squares) datasets. The dashed diagonal line represents perfect prediction (where predicted values equal actual values). The distribution of data points exhibits exceptional alignment with this reference line, indicating remarkable predictive accuracy across all subsets of data. Statistical performance metrics displayed in the upper left corner provide quantitative confirmation of the model's efficacy. The coefficient of determination ($R^2 = 0.9996$) approaches unity, demonstrating that 99.96% of the variance in the actual values is captured by the model's predictions. The Mean Absolute Percentage Error (MAPE) of overall data prediction is exceedingly low at 0.25%, further substantiating the precision of the predictions. The data range spans approximately 300 to 900 units on both axes, with consistent prediction quality maintained throughout this interval. Notably, the close correspondence between performance on training, validation, and test sets confirms the model's robust generalization capabilities without apparent overfitting. The absence of systematic deviations from the perfect prediction line suggests the model has successfully captured the underlying relationships in the data without significant bias. These results collectively indicate an exceptionally well-calibrated predictive model with performance characteristics suitable for high-precision applications.

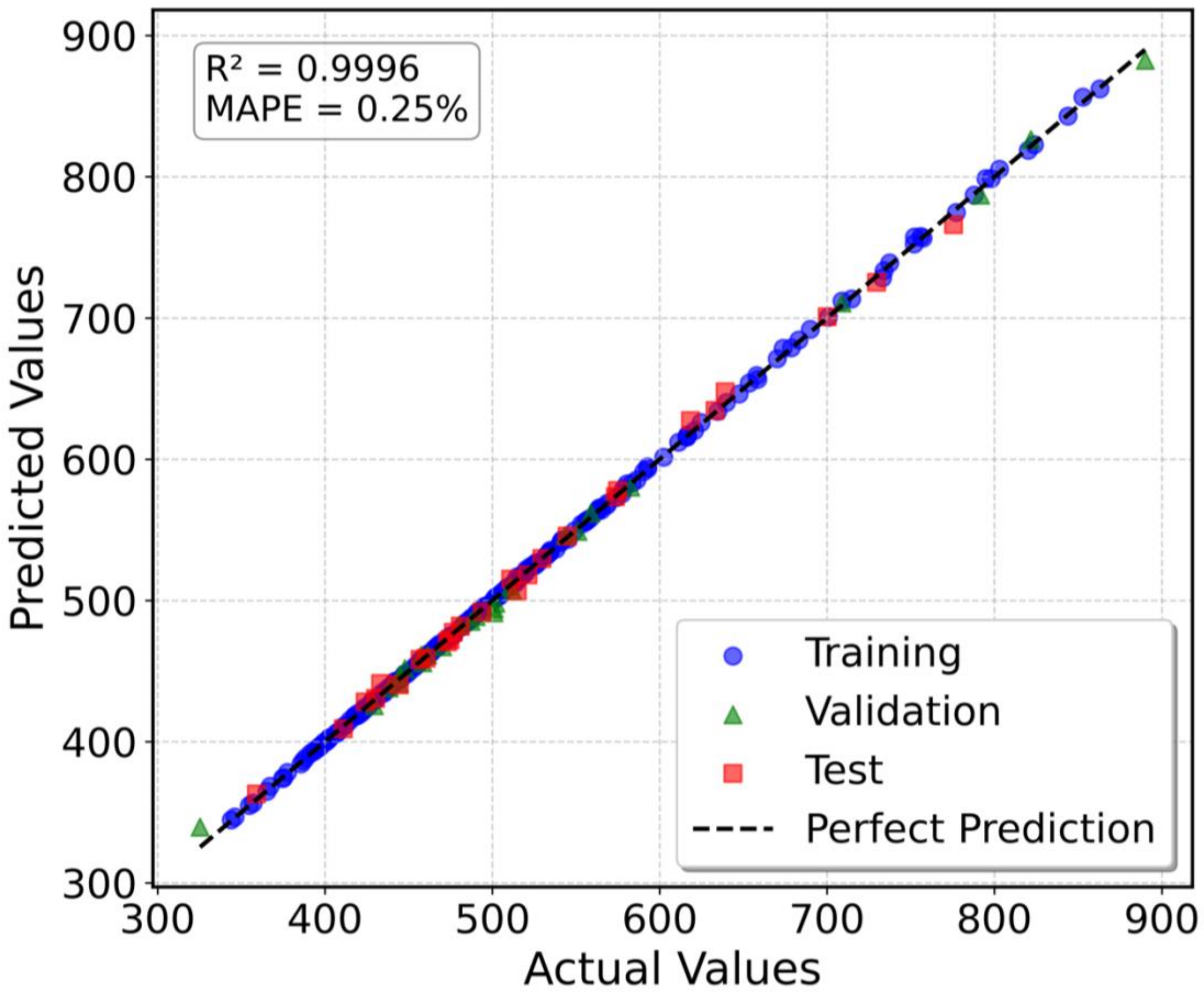

**Fig. 9** Predicted Values vs. Actual Values with Model Performance Metrics ($R^2$ = 0.9996, MAPE = 0.25%)

The model demonstrated exceptional performance, raising legitimate concerns regarding potential data leakage or intentional selection bias in the test set that might compromise the trustworthiness of the results. To address these concerns and evaluate model robustness, a comprehensive cross-validation was conducted using a 10-fold methodology, in which the dataset was partitioned into ten equal segments with each segment serving as a validation set once while the remaining data formed the training set. The training protocol was rigorously designed to run up to 500 epochs per fold, with early stopping criteria based on optimal validation loss performance rather than fixed iteration counts, thus preventing overfitting while ensuring sufficient model convergence. This approach represents standard practice in machine learning validation procedures and strengthens confidence in the generalizability of the findings. Cross-validation results yielded a Mean Absolute Percentage Error (MAPE) of 1.06% ± 0.30%, and $R^2$ = 0.9932 ± 0.0060, indicating remarkably consistent performance across all data partitions. Such low variance in error metrics strongly suggests that the model possesses robust generalization capabilities and is not merely memorizing training examples. Furthermore, these results indicate that the dataset size was appropriate for the model complexity, striking an effective balance between underfitting and overfitting. The combination of low error rates and minimal variation across folds provides compelling evidence that the model can reliably generalize to unseen data within the same distribution, although external validation on entirely independent datasets would further strengthen these conclusions.

### 4.1 Ablation study and benchmark

In this research, an ablation study was conducted to systematically evaluate model performance by analyzing individual components and their contributions to overall efficacy as shown in Table 3. It is evaluated with the test data same as the previous section at 10% of tap data. When our DATT architecture was stripped down to just a traditional multilayer perceptron with architecture [128,64,32,1], it demonstrated the highest error rate (9.66%), indicating that without attention mechanisms, the model has limited capacity to capture the inherent complexities of the dataset. This poor performance can be attributed to the MLP's inability to effectively model interdependencies between features and instances simultaneously, particularly in high-dimensional tabular data where relationships are often non-linear and context-dependent. We then examined the "Single feature attention" variant (0.74% MAPE), which utilized our DATT architecture but with the batch-level attention mechanism removed, retaining only feature-level attention. This modification allowed us to quantify the specific contribution of cross-instance learning. The substantial improvement over the base MLP (87% error reduction) demonstrates that feature-level attention significantly enhances the model's ability to identify and leverage relevant feature interactions by dynamically weighting their importance based on context.

For comparative benchmarking against state-of-the-art methodologies, we evaluated several leading approaches as shown in Table 3. Gradient boosting frameworks showed strong performance as standalone models, with Catboost and XGBoost achieving 2.52% and 1.83% MAPE, respectively. Their relative success can be attributed to their ensemble learning advantages and inherent robustness to non-linear relationships through sequential tree building and gradient optimization. However, these models still fall short in capturing complex feature dependencies that require more sophisticated attention mechanisms. TabPFN (Hollmann et al., 2025), the current state-of-the-art approach, exhibited performance of 0.82% MAPE, likely due to its prior-data fitted network approach that leverages meta-learning principles. TabPFN's ability to transfer knowledge across datasets gives it an edge over traditional boosting methods, but it lacks the explicit feature-instance interaction modeling that attention mechanisms provide. Our proposed complete Dual Attention Transformer (DATT) model significantly outperformed all existing methods, achieving superior performance with the lowest MAPE (0.63%), representing a decisive 23.2% relative improvement over the state-of-the-art TabPFN. This remarkable enhancement clearly establishes DATT as the new state-of-the-art, which can be attributed to its sophisticated dual attention architecture that simultaneously captures both feature-level and instance-level dependencies. The dual attention mechanism allows DATT to not only identify important features but also to recognize patterns across similar instances in the dataset, enabling a comprehensive understanding of the underlying data structure that previous methods cannot achieve. Furthermore, the synergistic effect of combining both attention types creates a multiplicative benefit that exceeds the sum of their individual contributions, as evidenced by the performance gap between the single feature attention variant and the complete DATT model.

**Table 3** Ablation study of the model

| Model | MAPE (%) of test data |
|---|---|
| MLP<br>[128,64,32,1] | 9.66 |
| Catboost | 2.52 |
| XGboost | 1.83 |
| TabPFN (Hollmann et al., 2025) | 0.82 |
| Single feature attention | 0.74 |
| **Proposed model: DATT** | **0.63** |

## 5. Explainable Model and Application

We utilized SHAP (SHapley Additive exPlanations) (Lundberg and Lee, 2017) analysis to interpret model predictions through global and local feature contributions. Global SHAP quantifies average feature contributions across the dataset, identifying key drivers of model decisions. Local SHAP provides instance-specific explanations, detailing how individual features affect predictions for particular data points, such as friction angle and cohesion prediction based on soil properties. Based on Shapley values from cooperative game theory, SHAP ensures fair attribution of feature contributions, satisfying properties of local accuracy, missingness, and consistency. This approach enhances interpretability of complex models, supporting both optimization and practical applications in geotechnical engineering.

SHAP (SHapley Additive exPlanations) analysis was employed to interpret the model's predictions by quantifying the contribution of each feature to both global and local decision-making processes. The fundamental equation for calculating SHAP values for a specific feature i is given by (Equation 17) :

$$\phi_i(v, x) = \sum_{S \subseteq N \setminus \{i\}} \frac{|S|!(|N| - |S| - 1)!}{|N|!} [v(S \cup \{i\}) - v(S)] \tag{17}$$

Here, $\phi_i$ represents the SHAP value for feature, $N$ is the complete set of features, $S$ represents any subset of features excluding i, $v$ is the prediction function and $x$ is the specific being explained. The term $\Delta_i(S) = v(S \cup \{i\}) - v(S)$ denotes the marginal contribution of feature i to the coalition S, while the coalition weight (Equation 18):

$$w(|S|) = \frac{|S|!(|N|-|S|-1)!}{|N|!} \quad (18)$$

It ensures fair attribution by accounting for the size of the coalition relative to the total feature set. For neural network models, the prediction function $v(S)$ is defined as the expected prediction given the subset $S$ of features(Equation 19):

$$v(S) = E[f(x)|x_S] \quad (19)$$

where $x_S$ denotes the instance with only features in $S$ activated, and the remaining features are masked or set to baseline values. This allows SHAP to decompose the model's prediction into an **additive feature attribution** (Equation 20):

$$f(x) = \phi_0 + \sum_{i=1}^{M} \phi_i \quad (20)$$

For a soil resistivity prediction model (Equation 21):

$$\text{Resistivity}(x) = \text{E}[\text{Resistivity}] + \sum_{i=1}^{M} \phi_i \quad (21)$$

Given the computational complexity of exact SHAP value calculation, the KernelExplainer approximates SHAP values using a weighted linear regression framework (Equation 22):

$$\phi_i \approx \sum_{z' \in Z} \frac{|Z|}{|Z'||Z|} \left[\hat{f}(h_x(z')) - \hat{f}\left(h_{x \setminus i}(z')\right)\right] \quad (22)$$

Here, $Z$ is a background dataset, $h_x$ maps simplified inputs (binary feature presence/absence) to the original feature space, and $\hat{f}$ is the trained model. The weights are derived from the proximity of background samples to the instance $x$, ensuring that local patterns are prioritized. The regression weights for the linear approximation are given by (Equation 23)::

$$w_i = \frac{|N|-1}{\binom{|N|-1}{|z_i|}|z_i|(|N|-|z_i|)} \quad (23)$$

where $|z_i|$ is the number of non-zero elements in each sample, balancing the influence of feature coalitions.

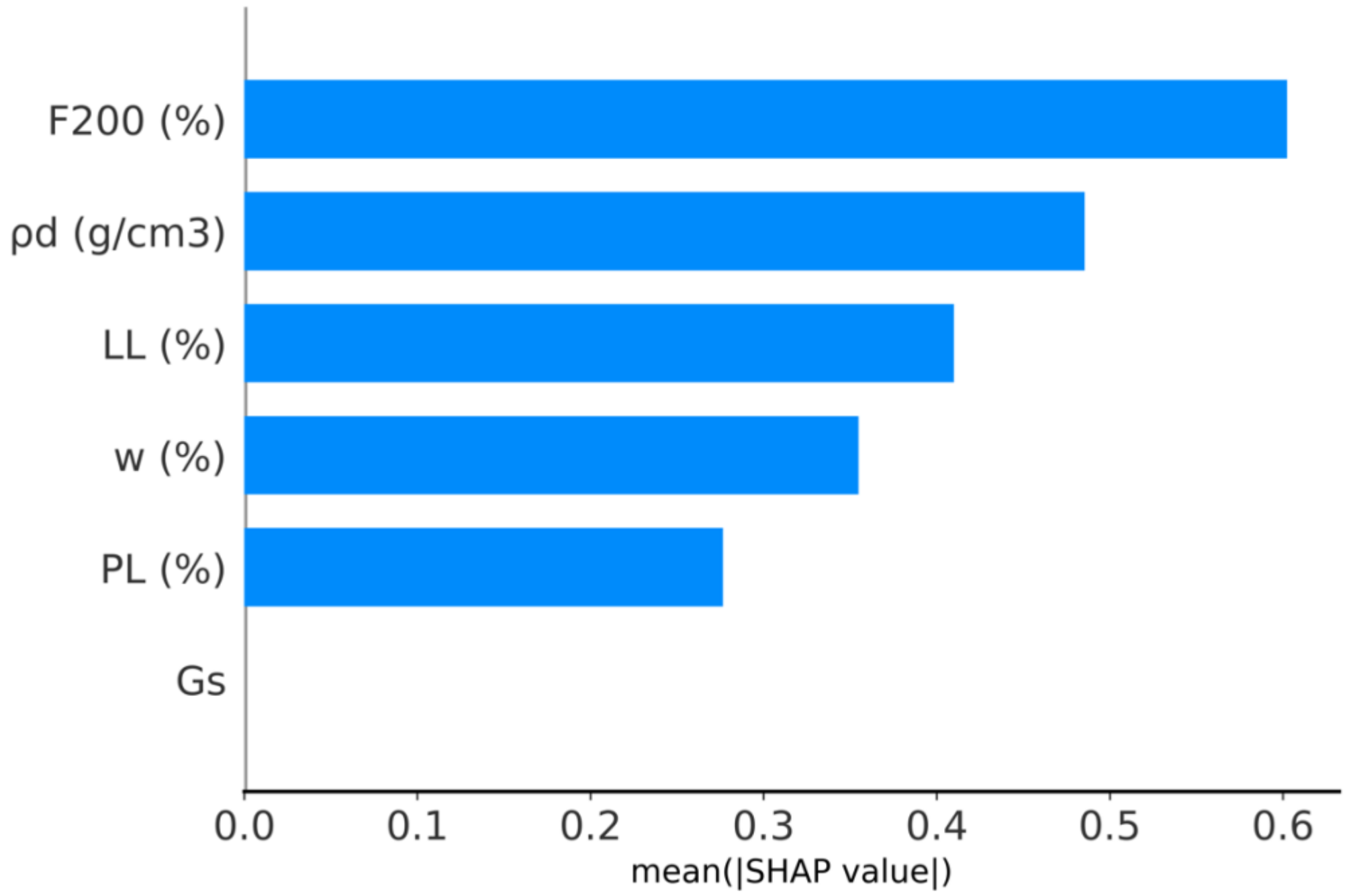

**Fig. 10** Mean Absolute SHAP Value Distribution of Geotechnical Engineering Parameters in Soil Resistivity Modeling

The bar chart in Fig. 10 quantifies parameter importance in soil resistivity prediction through mean absolute SHAP values. F200 (percent passing No. 200 sieve) exhibits highest importance (≈0.6), attributable to its representation of clay content which introduces surface conduction mechanisms alongside pore fluid pathways. Dry density (ρd) ranks second (≈0.5), controlling pore network architecture critical for ion mobility. Atterberg limits (LL, PL) demonstrate moderate influence, reflecting mineralogical composition effects on surface conductivity. Moisture content (w%) shows intermediate importance as the primary electrolytic medium. Specific gravity (Gs) exhibits negligible contribution, consistent with its limited electrochemical relevance. This hierarchical arrangement aligns with fundamental principles of soil electrical behavior where textural characteristics and moisture parameters govern the parallel conduction mechanisms (electrolytic and interfacial) that determine bulk resistivity properties in geomaterials.

The SHAP (SHapley Additive exPlanations) feature importance plot (Fig. 11) illustrates the quantitative impact of various soil properties on electrical resistivity predictions. The horizontal axis represents SHAP values, where negative values indicate decreased resistivity and positive values signify increased resistivity. Feature magnitude is denoted by color gradient from blue (low values) to pink (high values). Analysis of F200% (percent passing No. 200 sieve) reveals a multimodal distribution with high values (pink) predominantly clustered in both slightly negative and moderately positive SHAP regions, while lower values (blue) are concentrated in the far negative region. This bimodal pattern for high F200% values suggests that fine content influences resistivity through competing mechanisms depending on associated soil characteristics, which aligns with the anomalous findings discussed in the document where clean sand can sometimes exhibit lower resistivity than sand with fine content under specific conditions. Dry density (ρd g/cm$^3$) demonstrates a distinctive inverse relationship where higher values (pink) predominantly generate negative SHAP values, indicating that increased density generally decreases resistivity, likely due to enhanced pore connectivity facilitating ionic transport. Conversely, lower density values (blue) correspond to positive SHAP values, suggesting reduced connectivity increases resistivity. Liquid Limit (LL%) exhibits clustering of high values (pink) in the negative SHAP region with moderate-to-low values (purple-blue) in the positive region, indicating that higher plasticity typically correlates with decreased resistivity due to enhanced cation exchange capacity of clay minerals. These relationships collectively support modified Archie's Law formulations (Shah and Singh, 2005; Waxman and Smits, 1968) for heterogeneous soil systems, where resistivity emerges from complex interactions between porosity, saturation degree, pore fluid properties, and surface conduction mechanisms at solid-liquid interfaces.

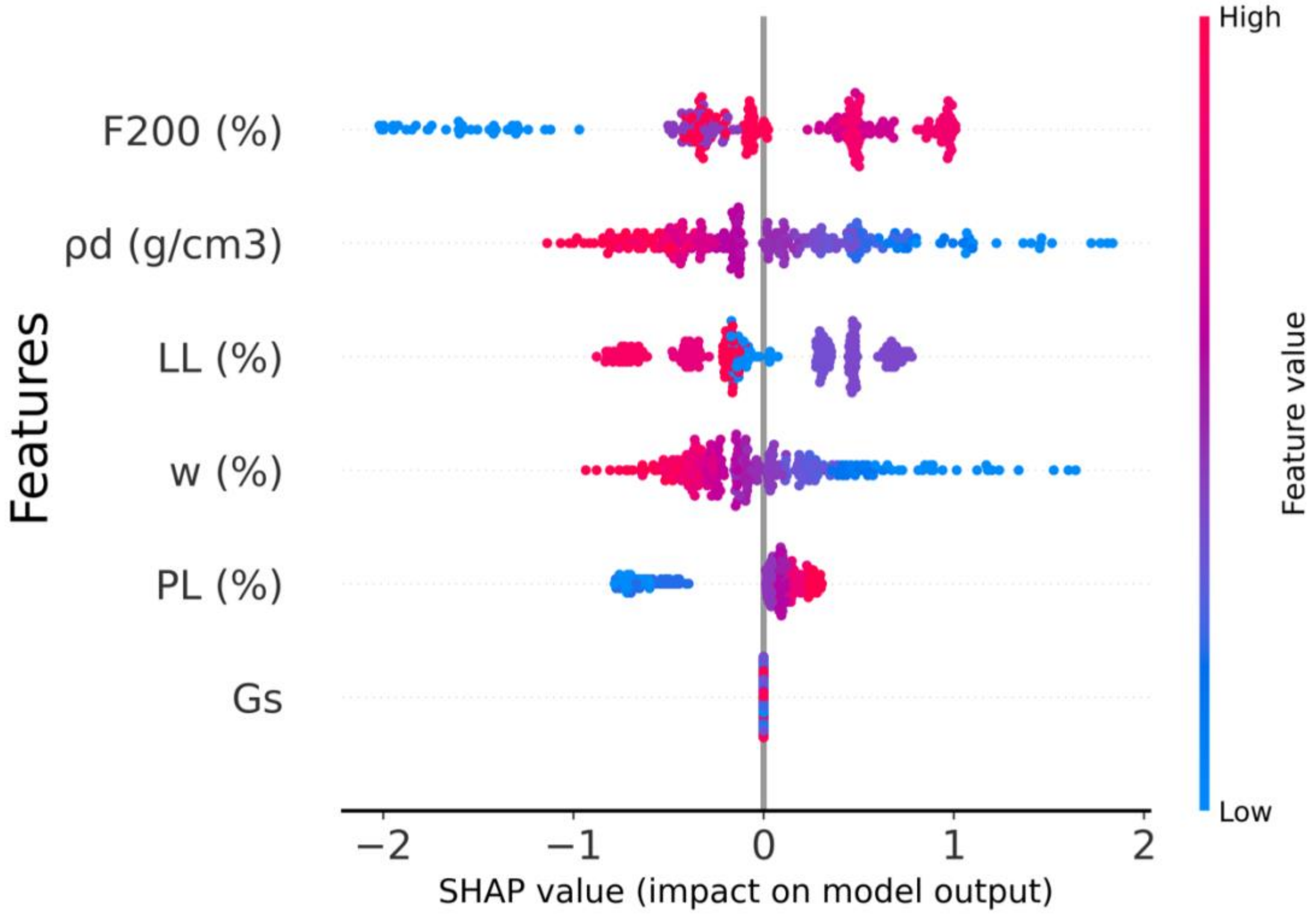

**Fig. 11** SHAP Value Distribution of Geotechnical Engineering Parameters: Impact Magnitude and Directionality on Soil Resistivity Model Predictions

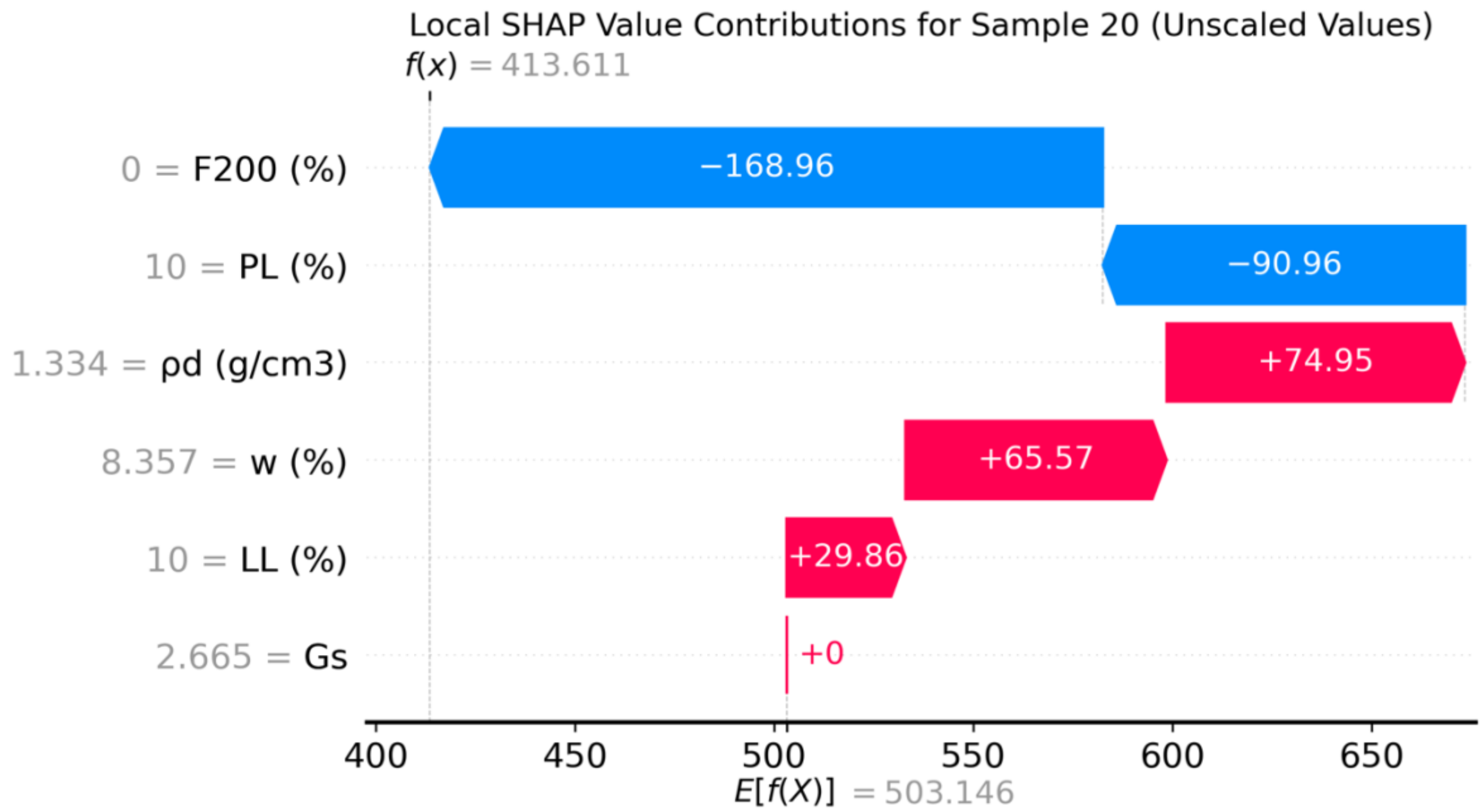

**Fig. 12** Local SHAP Value Contributions for Soil Resistivity Prediction: Quantitative Analysis of Feature Impacts on Model Output

The main application of this study extends beyond merely predicting soil resistivity suitable for establishing substations. It also illuminates feature importance through local SHAP (SHapley Additive exPlanations) values, as demonstrated in Figure 12. This waterfall plot quantifies how each soil parameter contributes to the prediction outcome, with a base value E[f(X)] of 503.146 and final prediction f(x) of 413.611. The visualization reveals that F200 (%) and PL (%) exert substantial negative influences of -168.96 and -90.96 units respectively, indicating these parameters decrease the predicted soil resistivity. This aligns with established soil physics principles, as finer particles typically increase water retention and ionic mobility, thereby reducing resistivity. Soil parameter interactions become more complex when examining the positive contributors: ρd (g/cm$^3$) and w (%) contribute positive effects of +74.95 and +65.57 units respectively, increasing the predicted value. The positive contribution of dry density can be attributed to tighter particle packing, which reduces pore connectivity and restricts ionic movement, while the counterintuitive positive influence of water content suggests complex interactions within this specific soil composition. LL (%) shows a moderate positive contribution of +29.86 units, potentially reflecting the soil's capacity to bind water molecules rather than allowing free ion movement, whereas Gs demonstrates negligible impact (+0). This comprehensive SHAP analysis provides critical insights into the electro-physical mechanisms governing soil resistivity, offering a quantitative framework for understanding how individual soil properties modulate current flow through the earth. Users can trace exactly how the model arrived at each prediction, identifying which soil properties had the greatest influence and in what direction. This transparency enables engineers to not only trust the model's output but also to understand the underlying physical relationships between soil composition and electrical properties—knowledge essential for electrical grounding system design and corrosion protection strategies in substation construction.

## 5. Application

This study developed a web-based application to predict soil resistivity based on varying feature values as shown in Fig. 13. The application implements the deep learning model and is publicly accessible through the Hugging Face cloud system, providing engineers with a convenient and accessible tool for soil resistivity prediction. The interface allows users to input soil parameters including bulk density (ρd), water content (w), particle size distribution (F200), specific gravity (Gs), liquid limit (LL), and plasticity limit (PL). Upon submission of these parameters, the model calculates the predicted soil resistivity value and displays both the numerical result and a feature importance graph illustrating the contribution of each parameter to the final prediction. This visualization helps engineers understand the relative influence of different soil properties on resistivity measurements, facilitating more informed geotechnical decision-making in the field. The application serves as a real-time tool for engineers to estimate soil resistivity values directly from field or laboratory measurements, eliminating the need for time-consuming traditional resistivity testing methods and enabling rapid assessment of site conditions during preliminary investigations or design phases.

Future development of this application could include integration with GIS systems to allow for spatial mapping of predicted resistivity values across project sites. Additionally, the incorporation of a database functionality would enable users to store historical measurements and predictions, facilitating trend analysis and improving prediction accuracy through continuous model refinement. The open-ended nature of this platform presents opportunities for expansion into related geotechnical applications, such as predicting thermal conductivity, hydraulic conductivity, or corrosion potential. Mobile device compatibility could be enhanced to support field engineers conducting in-situ assessments, potentially incorporating direct input from field testing devices. Furthermore, the addition of an uncertainty quantification feature would provide engineers with confidence intervals for predictions, supporting more robust risk assessment in geotechnical design projects.

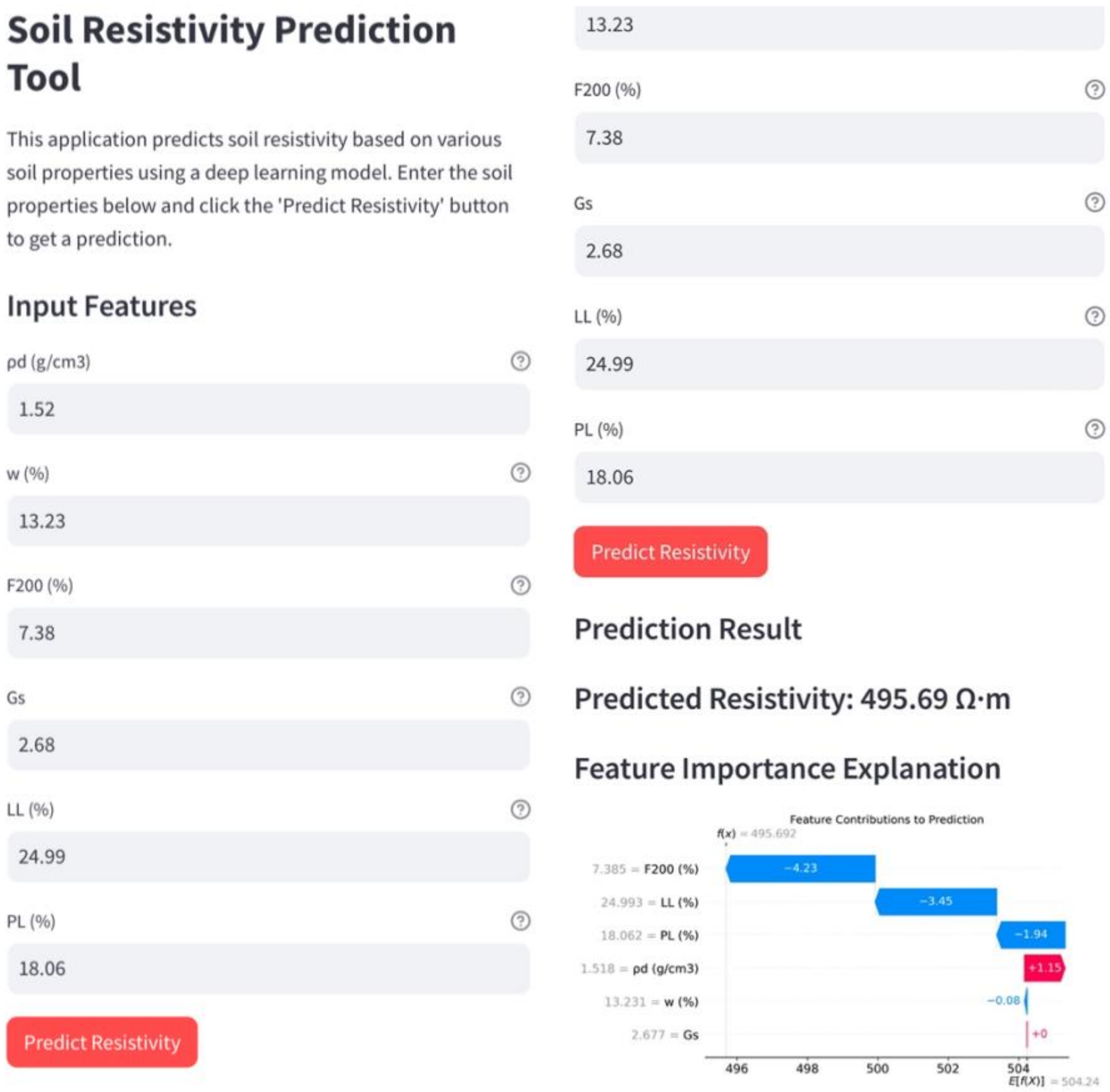

**Fig. 13** The web-based application for prediction the soil resistivity and local SHAP value of the parameter's contribution
< https://huggingface.co/spaces/Sompote/SoilResistivity>

## 6. Discussion

The superior performance of the proposed Dual-Attention Tabular Transformer (DATT) architecture in soil resistivity prediction can be attributed to several key factors that differentiate it from existing methodologies. First, the integration of orthogonal attention mechanisms—operating across both features and samples—enables the model to simultaneously capture complex dependencies between soil parameters and contextual patterns across the dataset. This dual-attention approach addresses the fundamental limitation of traditional neural networks and gradient boosting methods that process features either sequentially or hierarchically without considering their global interactions.

The embedding layer's transformation of raw features into higher-dimensional representations (optimized at 36 dimensions) provides the model with enhanced capacity to encode non-linear relationships between soil properties and electrical resistivity. This approach shares conceptual similarities with token embedding in large language models, where dimensional expansion facilitates more nuanced pattern recognition. The performance improvement of 23.2% over TabPFN and 14.9% over single-feature attention validates our architectural hypothesis that two-dimensional attention captures essential physical mechanisms governing soil electrical properties more effectively than one-dimensional approaches.

SHAP analysis revealed that fine particle content (F200%) exerts the most significant influence on resistivity predictions, followed by dry density (ρd) and Atterberg limits (LL, PL). This finding aligns with established geophysical principles where clay content affects both electrolytic pathways through pore water and surface conduction mechanisms at particle interfaces. The counterintuitive positive correlation between water content and resistivity in certain samples, as visualized in the SHAP plots, suggests complex interactions between moisture and mineralogical composition that simple empirical models would fail to capture. Our model successfully identified these non-trivial relationships without requiring predefined functional forms, demonstrating its ability to learn directly from data patterns.

The hyperparameter optimization via Particle Swarm Optimization (PSO) represents a methodological advancement over manual tuning approaches commonly employed in geotechnical engineering applications. By systematically exploring the parameter space, we identified optimal configurations that balance model complexity and generalization capability. The rapid convergence observed in the loss curves (stabilizing within 50 epochs) further indicates the efficiency of our optimization strategy in identifying robust parameter settings.

Despite the promising results, several limitations warrant consideration for future research. First, the dataset composition primarily encompasses lateritic soils and fine sands, potentially limiting the model's generalizability to other soil types such as organic soils, expansive clays, or coarse-grained materials with significantly different electrical properties. The constrained geographic origin of samples (primarily from Thailand) introduces potential regional bias, as soil mineralogy and formation history vary considerably across different geological contexts. Second, while our model achieved exceptional accuracy on laboratory-prepared specimens, field conditions introduce additional variables including heterogeneity, anisotropy, and temporal variations due to seasonal moisture fluctuations—factors not fully captured in our controlled experimental setup. The model's performance under such variable conditions requires further validation through comprehensive field testing across diverse geological environments. Third, the current implementation focuses exclusively on soil resistivity prediction without considering related electrochemical properties such as chargeability or frequency-dependent impedance, which provide complementary information relevant to corrosion potential and earthing system design. Extending the model to predict multiple electrical parameters simultaneously would enhance its practical utility for comprehensive substation design.

The transformer-based architecture implemented in this study demonstrates remarkable computational efficiency compared to conventional transformer models. Despite being built on transformer principles, our model achieves significant parameter reduction with only 2.352k parameters, requiring exceptionally low computational resources at 0.000058 GFLOPs for training, which enables model training on standard CPU-only systems without specialized hardware acceleration. Performance testing on a MacBook Pro with a 2.3 GHz Intel Core i9 processor yielded inference times of less than one second, highlighting the model's suitability for real-time applications, while the Particle Swarm Optimization (PSO) component completes within 10 minutes on standard CPU hardware, eliminating the need for GPU acceleration during the optimization phase. These efficiency achievements position our model as particularly valuable for resource-constrained environments and suggest promising directions for democratizing access to transformer-based models across diverse research communities and application domains where computational constraints might otherwise preclude the deployment of transformer architectures.

From a methodological perspective, the computational complexity of dual-attention mechanisms represents a practical constraint for deployment on resource-limited field devices. Although inference time remains reasonable for individual predictions (suitable for engineering decision support), real-time spatial mapping applications would benefit from model optimization or distillation techniques to reduce computational requirements. Finally, while SHAP analysis provides valuable insights into feature importance, it quantifies correlative relationships rather than establishing causal mechanisms between soil properties and electrical behavior. Integrating domain-specific physical constraints or prior knowledge into the model architecture could further enhance interpretability and ensure predictions remain physically plausible across the entire parameter space, particularly for extrapolation beyond the training distribution.

## 7. Conclusion

This research introduced a novel Dual-Attention Tabular Transformer (DATT) architecture for predicting soil electrical resistivity, addressing a critical need in high-voltage substation construction. Our architecture advances the state-of-the-art by implementing simultaneous attention mechanisms along two orthogonal dimensions - feature attention and batch attention - enabling the model to capture complex dependencies both within individual soil samples and across the dataset distribution. Unlike previous approaches, DATT incorporates an embedding layer that projects each feature into a higher-dimensional space (optimized to 36 dimensions), followed by multi-head self-attention mechanisms (4 heads) that operate iteratively to extract hierarchical patterns in the data.

Comparative analysis demonstrated DATT's significant performance advantages over the state-of-the-art TabPFN model, achieving a 23.2% relative improvement in prediction accuracy with a Mean Absolute Percentage Error (MAPE) of 0.63% compared to TabPFN's 0.82%. This enhancement stems from DATT's sophisticated dual-attention mechanism that more effectively models the non-linear relationships between soil parameters and resistivity than TabPFN's prior-data fitted approach. While TabPFN leverages meta-learning and pre-training on synthetic data, our model's specialized architecture captures domain-specific relationships directly from the soil data, resulting in superior generalization to unseen samples.

The integration of SHAP analysis provided critical insights into the model's decision-making process, revealing that fine particle content (F200%) and dry density (ρd) exert the most significant influence on soil resistivity predictions. This explainability component addresses the "black box" limitations of previous approaches, offering transparent interpretations of parameter interactions that determine electrical properties.

Our web-based implementation demonstrates DATT's practical utility as a decision support tool for organizations like the Electricity Generating Authority of Thailand, enabling accurate prediction of soil resistivity from standard geotechnical parameters. By bridging the gap between geotechnical engineering requirements and electrical performance criteria, this research establishes quantifiable correlations that enhance construction efficiency, minimize remediation costs, and improve safety compliance in high-voltage substation implementation.

## 8. References

Alsharari, B., Olenko, A., Abuel-Naga, H., 2020. Modeling of electrical resistivity of soil based on geotechnical properties. Expert Systems with Applications 141, 112966. https://doi.org/10.1016/j.eswa.2019.112966

Alzo'ubi, A.K., Ibrahim, F., 2018. Predicting Loading–Unloading Pile Static Load Test Curves by Using Artificial Neural Networks. Geotechnical and Geological Engineering 37, 1311–1330.

ASTM International, 2020. Standard Test Method for Measurement of Soil Resistivity Using the Wenner Four-Electrode Method (No. ASTM G57-20). ASTM International, West Conshohocken, PA. https://doi.org/10.1520/G0057-20

Bai, W., Kong, L., Guo, A., 2013. Effects of physical properties on electrical conductivity of compacted lateritic soil. Journal of Rock Mechanics and Geotechnical Engineering 5, 406–411. https://doi.org/10.1016/j.jrmge.2013.07.003

Bazaluk, B., Hamdan, M., Ghaleb, M., Gismalla, M.S.M., da Silva, F.S.C., Batista, D.M., 2024. Towards a Transformer-Based Pre-trained Model for IoT Traffic Classification. https://doi.org/10.48550/ARXIV.2407.19051

Bian, L., Qin, X., Zhang, C., Guo, P., Wu, H., 2023. Application, interpretability and prediction of machine learning method combined with LSTM and LightGBM-a case study for runoff simulation in an arid area. Journal of Hydrology 625, 130091. https://doi.org/10.1016/j.jhydrol.2023.130091

Cardoso, R., Dias, A.S., 2017. Study of the electrical resistivity of compacted kaolin based on water potential. Engineering Geology 226, 1–11. https://doi.org/10.1016/j.enggeo.2017.04.007

Chantachot, T., Kongkitkul, W., Youwai, S., Jongpradist, P., 2016. Behaviours of geosynthetic-reinforced asphalt pavements investigated by laboratory physical model tests on a pavement structure. Transportation Geotechnics 8, 103–118. https://doi.org/10.1016/j.trgeo.2016.03.004

Chehreh Chelgani, S., Homafar, A., Nasiri, H., Rezaei Laksar, M., 2024. CatBoost-SHAP for modeling industrial operational flotation variables – A "conscious lab" approach. Minerals Engineering 213, 108754. https://doi.org/10.1016/j.mineng.2024.108754

Chen, T., Guestrin, C., 2016. XGBoost: A Scalable Tree Boosting System, in: Proceedings of the 22nd ACM SIGKDD International Conference on Knowledge Discovery and Data Mining, KDD '16. ACM, New York, NY, USA, pp. 785–794. https://doi.org/10.1145/2939672.2939785

Cholakov, R., Kolev, T., 2022. The GatedTabTransformer. An enhanced deep learning architecture for tabular modeling. https://doi.org/10.48550/ARXIV.2201.00199

Dararat, S., Kongkitkul, W., Arangjelovski, G., Ling, H.I., 2021. Estimation of stress state-dependent elastic modulus of pavement structure materials using one-dimensional loading test. Road Materials and Pavement Design 22, 245–267. https://doi.org/10.1080/14680629.2019.1620119

ForouzeshNejad, A.A., Arabikhan, F., Aheleroff, S., 2024. Optimizing Project Time and Cost Prediction Using a Hybrid XGBoost and Simulated Annealing Algorithm. Machines 12, 867. https://doi.org/10.3390/machines12120867

Hollmann, N., Müller, S., Purucker, L., Krishnakumar, A., Körfer, M., Hoo, S.B., Schirrmeister, R.T., Hutter, F., 2025. Accurate predictions on small data with a tabular foundation model. Nature 637, 319–326. https://doi.org/10.1038/s41586-024-08328-6

Huang, X., Khetan, A., Cvitkovic, M., Karnin, Z., 2020. TabTransformer: Tabular Data Modeling Using Contextual Embeddings. https://doi.org/10.48550/ARXIV.2012.06678

IEEE Power and Energy Society, 2012. IEEE Guide for Measuring Earth Resistivity, Ground Impedance, and Earth Surface Potentials of a Grounding System (Standard No. IEEE Std 81-2012). IEEE, New York, NY, USA.

Jariyatatsakorn, K., Kongkitkul, W., Tatsuoka, F., 2024. Prediction of creep strain from stress relaxation of sand in shear. Soils and Foundations 64, 101472. https://doi.org/10.1016/j.sandf.2024.101472

Ke, G., Meng, Q., Finley, T., Wang, T., Chen, W., Ma, W., Ye, Q., Liu, T.-Y., 2017. LightGBM: A Highly Efficient Gradient Boosting Decision Tree, in: Neural Information Processing Systems.

Kennedy, J., Eberhart, R., 1995. Particle swarm optimization, in: Proceedings of ICNN'95 - International Conference on Neural Networks. Presented at the ICNN'95 - International Conference on Neural Networks, IEEE, Perth, WA, Australia, pp. 1942–1948. https://doi.org/10.1109/ICNN.1995.488968

Kingma, D.P., Ba, J., 2014. Adam: A Method for Stochastic Optimization. https://doi.org/10.48550/ARXIV.1412.6980

Kongkitkul, W., Punthutaecha, K., Youwai, S., Jongpradist, P., Moryadee, S., Posribink, T., Bamrungwong, C., Hirakawa, D., 2011. Simple Dynamic Hammer for Evaluation of Physical Conditions of Pavement Structures. Transportation Research Record: Journal of the Transportation Research Board 2204, 35–44. https://doi.org/10.3141/2204-05

Lundberg, S.M., Lee, S.-I., 2017. A Unified Approach to Interpreting Model Predictions, in: Guyon, I., Luxburg, U.V., Bengio, S., Wallach, H., Fergus, R., Vishwanathan, S., Garnett, R. (Eds.), Advances in Neural Information Processing Systems. Curran Associates, Inc., pp. 4765–4774.

Ozcep, F., Yıldırım, E., Tezel, O., Asci, M., Karabulut, S., n.d. Correlation between electrical resistivity and soil-water content based artificial intelligent techniques. Int. J. Phys. Sci.

Prokhorenkova, L., Gusev, G., Vorobev, A., Dorogush, A.V., Gulin, A., 2018. CatBoost: unbiased boosting with categorical features. arXiv preprint arXiv:1706.09516.

Riani, M., Atkinson, A.C., Corbellini, A., 2023. Automatic robust Box–Cox and extended Yeo–Johnson transformations in regression. Stat Methods Appl 32, 75–102. https://doi.org/10.1007/s10260-022-00640-7

Sangprasat, K., Puttiwongrak, A., Inazumi, S., 2024. Comprehensive analysis of correlations between soil electrical resistivity and index geotechnical properties. Results in Engineering 23, 102696. https://doi.org/10.1016/j.rineng.2024.102696

Shah, P., Singh, D., 2005. Generalized Archie's Law for Estimation of Soil Electrical Conductivity. Journal of ASTM International 2, 1–20. https://doi.org/10.1520/JAI13087

Sinha, B.B., Ahsan, M., Dhanalakshmi, R., 2023. LightGBM empowered by whale optimization for thyroid disease detection. Int. j. inf. tecnol. 15, 2053–2062. https://doi.org/10.1007/s41870-023-01261-3

Tang, S., 2024. The box office prediction model based on the optimized XGBoost algorithm in the context of film marketing and distribution. PLoS ONE 19, e0309227. https://doi.org/10.1371/journal.pone.0309227

Truong, V.-H., Tangaramvong, S., Papazafeiropoulos, G., 2024. An efficient LightGBM-based differential evolution method for nonlinear inelastic truss optimization. Expert Systems with Applications 237, 121530. https://doi.org/10.1016/j.eswa.2023.121530

Vaswani, A., Shazeer, N., Parmar, N., Uszkoreit, J., Jones, L., Gomez, A.N., Kaiser, L., Polosukhin, I., 2017. Attention Is All You Need. https://doi.org/10.48550/ARXIV.1706.03762

Vyas, T.K., 2024. Deep Learning with Tabular Data: A Self-supervised Approach. https://doi.org/10.48550/ARXIV.2401.15238

Waxman, M.H., Smits, L.J.M., 1968. Electrical Conductivities in Oil-Bearing Shaly Sands. Society of Petroleum Engineers Journal 8, 107–122. https://doi.org/10.2118/1863-A

Youwai, S., Wongsala, K., 2024. Predicting the Friction Angle of Bangkok Sand Using State Parameter and Neural Network. Geotech Geol Eng. https://doi.org/10.1007/s10706-024-02873-7

Zhou, J., Wang, S., Lai, K.K., 2024. An Efficient and Interpretable XGBoost Framework for Housing Price Prediction: Enhancements in Accuracy and Computational Performance. International Journal of Computer Engineering in Research Trends 11, 1234–1245. https://doi.org/10.1016/j.ijcet.2024.09.012